\documentclass[a4paper,fleqn]{cas-sc}

\usepackage[authoryear,longnamesfirst,square]{natbib}
\usepackage{subcaption}
\usepackage{amsthm}

\newtheorem{Lemma}{Lemma}
\newtheorem{Theorem}{Theorem}
\newtheorem*{Proof}{Proof}

\usepackage{xcolor}

\def\tsc#1{\csdef{#1}{\textsc{\lowercase{#1}}\xspace}}
\tsc{WGM}
\tsc{QE}
\tsc{EP}
\tsc{PMS}
\tsc{BEC}
\tsc{DE}

\begin{document}
\let\WriteBookmarks\relax
\def\floatpagepagefraction{1}
\def\textpagefraction{.001}
\shorttitle{Epidemic Transmission Modelling on the Birth-death Evolving Network with Indirect Contacts}
\shortauthors{Minyu Feng et~al.}

\title [mode = title]{Epidemic Transmission Modelling on the Birth-death Evolving Network with Indirect Contacts}                      

\cortext[1]{Corresponding author. Yuhan Li}



\author[1]{Minyu Feng}[orcid=0000-0001-6772-3017]
\fnmark[1]
\ead{myfeng@swu.edu.cn}

\affiliation[1]{organization={College of Artificial Intelligence, Southwest University, Chongqing, 400715, P. R. China.}}

\author[2]{Yuhan Li}[orcid=0000-0001-6445-2303]
\fnmark[2]
\cormark[1]
\ead{yuhan.li@qmul.ac.uk}

\affiliation[2]{organization={School
of Mathematical Sciences, Queen Mary University of London, London, E1 4NS, U.K.}}

\author[3,4]{J\"{u}rgen Kurths}
\fnmark[3]
\ead{Juergen.Kurths@pik-potsdam.de}
\affiliation[3]{organization={Department of Complexity Science, Potsdam Institute for Climate Impact Research, Potsdam, 14437, Germany}}

\affiliation[3]{organization={Institute of Physics, Humboldt
University of Berlin},
                city={Berlin},
                postcode={12489}, 
                country={Germany}}

\begin{abstract}
Epidemic modelling on complex networks has been studied intensively all the time. The majority of relative research assumes that the time scale of the underlying network evolution is much larger compared to the propagation dynamics on it, while the co-evolution of epidemics and networks needs exploring further. In this paper, we investigate how our recently proposed birth-death evolving network impacts the Susceptible-Infected-Recovered-Susceptible (SIRS) epidemic process. Our evolving network considers the increase and the heritable deletion of nodes, which enables to depicting individual behaviors during an epidemic, e.g., population migration and indirect contacts. To model the above processes, we construct a Markovian queueing network and perform analyses for the variation of population size of different epidemic states. In simulations, we reveal how the population migration and indirect contacts caused by our network dynamic properties influence the population sizes of each epidemic state, and find that newly-created indirect contacts facilitate epidemic transmission.
\end{abstract}



\begin{keywords}
epidemic model\sep 
evolving network\sep 
indirect contacts\sep 
Markovian queueing network\sep
infected population size
\end{keywords}



\maketitle

\section{Introduction}
Studying infectious diseases spreading on networks is of great significance to predicting and controlling epidemics in real life.
Epidemic transmissions were first studied via the susceptible-infected-recovered (SIR) compartmental model proposed by Kermack and Mckendrick \cite{kermack1927contribution}, laying a foundation for extended models such as Susceptible-Exposed-Infected-Recovered (SEIR) model \cite{li1995global}, Susceptible-Infected-Recovered-Susceptible (SIRS) model \cite{ghosh1995sirs}, etc. These compartmental models assume a homogeneous mix of population and contacts. Later, there were significant progresses in epidemiology due to the emergence of network science, which provided a useful framework for studying the dynamics of epidemic spreading. Network structures can concretely describe contact patterns of a population, e.g., the heterogeneity of the connections among people \cite{albert2002statistical}, which makes modelling more accurate.

As a lot of network models were proposed, the dynamics of epidemic spreading were investigated on various kinds of networks. After the Barab\'{a}si-Albert (BA) scale-free network was proposed \cite{barabasi1999emergence}, the epidemic threshold was studied on scale-free networks, which showed that the threshold inclines diminish on these heterogeneous networks \cite{pastor2001epidemic}. Recently, epidemic processes were studied based on multiplex networks \cite{wang2020epidemic} \cite{an2024coupled} \cite{luo2023modeling}, and higher-order networks \cite{chen2023composite} \cite{majhi2022dynamics} \cite{cisneros2021multigroup} \cite{fan2022epidemics} \cite{yuan2025impacts}.

In addition to static network structure, various dynamic network models emerged to describe dynamic complex systems in real life, and this dynamic property has a great impact on epidemic dynamics on networks. Temporal networks were established and the impact of time-reversed characteristics on the epidemic threshold was analyzed \cite{zhang2017spectral}. Stability analyses of epidemic processes on evolving networks were performed \cite{pare2017epidemic}. Epidemic spreading process on the activity-driven network \cite{perra2012activity} was investigated \cite{zino2016continuous} \cite{nadini2018epidemic}. Except for utilizing the active and inactive nodes and edges to describe dynamical networks, evolving networks were also constructed considering the growth and decrease of nodes. An evolving network was proposed with a removal of nodes and edges by probabilities at each time step as well as the growth mechanism identical BA model \cite{moore2006exact}. Further, the birth-death process was utilized to construct evolving networks in continuous time where the growth and deletion of nodes follow Poisson processes \cite{feng2016evolving}\cite{zeng2023temporal}, and another model regarded the decrease of nodes as the mortality of individuals induced by diseases \cite{demirel2017dynamics}. Besides, adaptive networks were proposed where individuals avoid contacting infected ones by rewiring their connections \cite{gross2006epidemic}. 


Apart from studies on the impact of network structures and properties on epidemic spreading, some important factors in real systems, i.e. individuals' behaviors in an epidemic were taken into consideration \cite{wang2016statistical}. Metapopulation models were proposed to describe the mobility of population in epidemics \cite{pastor2015epidemic} \cite{li2022network}, where nodes indicate areas and edges indicate traffic flows. In another model, there were logistic equations in each node to describe epidemic dynamics in an area \cite{gong2018epidemic}. Spatial travel and social factors were further considered based on the metapopulation model \cite{duan2022modeling}, and coupled propagation of awareness were studied \cite{gao2022epidemic}. Queueing systems were applied to describe the migration of population during epidemic spreading on networks where each epidemic state is regarded as a service center and individuals are customers \cite{li2021protection}. Epidemic processes are also investigated on higher-order networks where a link holds more than two nodes \cite{fan2022epidemics}, and the impact of personal resources on epidemic spreading was studied on higher-order networks \cite{sun2022diffusion}. Individuals' other behaviors and awareness \cite{paarporn2017networked}, e.g. self-protection \cite{lee2021epidemic} and self-quarantine \cite{wang2021impact}\cite{abbasi2024switched}, and lockdown \cite{gosak2021community} were also investigated in epidemic spreading processes.

Previous work contributes a lot to modelling epidemic spreading on networks. Underlying networks which represent the population contact structure may have an significant impact on the epidemic processes. It is essential to investigate the co-evolution of networks and epidemic spreading on the same time scale. We previously established a birth-death evolving network model with a heritable disconnection mechanism, as well as the growth and deletion of nodes \cite{feng2022heritable}. While the previous work \cite{feng2022heritable} focused on modeling the evolution of networks themselves, in this paper, we investigate dynamics of epidemic transmission on this specific evolving network. We interpret the network evolution in the perspective of individuals' behaviours. In detail, we regard the growth and deletion of nodes in a network as the inflow and outflow of population in a region. We regard the heritable disconnection mechanism as indirect contact effects, revealing practical implications of the evolution of our network within the context of epidemic spreading.

The organization of the paper is as follows: In Sec. \ref{sec:II}, we present the birth-death evolving network model with heritable deletions, model epidemic spreading on our proposed network, and perform analyses for our epidemic model. In Sec. \ref{sec:IV}, we simulate the epidemic spreading process with different parameters and demonstrate the influence of indirect contact on the population size of each epidemic state. Conclusions and future work are given in Sec. \ref{sec:V}.

\section{Epidemic Spreading on Proposed Evolving Networks}\label{sec:II}
There are a number of dynamic systems featured with network structures in real life. For instance, in a population contact network of an area where nodes represent people, there are people coming and leaving the area randomly. These natural random events can be described as Poisson processes since their occurrences within a unit of time follow Poisson distributions. Hence, we assume that the growth and decrease of nodes in our evolving network follow Poisson processes. In this section, we first display our proposed evolving network model, then study epidemic transmissions on the underlying evolving network, and finally perform theoretical analyses for our epidemic model.
\subsection{Evolving Network Model With Heritable Disconnections}\label{sec:II1}
In this sub-section, we primarily introduce our proposed evolving network model and heritable disconnection mechanisms for the evolving network \cite{feng2022heritable}.

The evolution of the proposed network includes both the growth and the decrease of the network. The input of nodes implies network growth and the output of nodes indicates the network decrease. The input and output of nodes are regarded as stochastic processes, and time is continuous in our model. In particular, the input and output of nodes is a birth and death process, both of which follow the Poisson process with a rate $\lambda$ and $\mu$ respectively. This indicates that new nodes are generated and join the network randomly at a rate of Poisson parameter $\lambda$, and nodes leave the network at a rate of Poisson parameter $\mu$. In this case, there is a lifespan for each node in the network, which follows the exponential distribution with a parameter $\mu$. The lifespan of a node follows
\begin{equation}
P(t)=\left\{
\begin{array}{rcl}
1-e^{-\mu t}, & & {t\geq0}\\
0, & & {\text{otherwise}}
\end{array}
\right.
\end{equation}
The network evolves driven by the input and output of nodes as described above.

We next introduce the connection and disconnection mechanism of the evolving network. As for the connection mechanism, new nodes come with $m$ edges, and each edge connects to existing nodes in the network by the preferential attachment in BA scale-free network model. The probability of a new edge connecting to vertex $i$ in the network $\Pi_i$ is
\begin{equation}\label{pa}
\Pi_i=\frac{k_i}{\sum_{i}k_i}
\end{equation}
where $k_i$ is the current degree of vertex $i$.

As for the disconnection of nodes leaving the network, we proposed a heritable disconnection mechanism. Different from brutal deletions (deleting the leaving nodes and all their connections), we take the inheritance of connections which occurs in many real inheritance networks into consideration. For example, in a social network where nodes indicate people and edges indicate connections, if a person died, their closest family will inherit their social ties. Similarly, during an epidemic, when an individual leaves an area, they may have an influence on the contact structure of the population by introducing another individual $i$ to their neighbors by inheriting their connections to the individual $i$. Ttheir bridges individual $i$ connected to their neighbors indirectly, which we call indirect contact. Notably, we assume that similar nodes are geographically close in our network. Nodes are more likely to contact with other similar nodes than with unsimilar ones in the network. This assumption aligns with empirical evidence, as individuals who share similar traits such as occupation or social status are more likely to live or work in close proximity. Besides, literature supports the notation that physical contact networks exhibit assortative mixing, where nodes preferentially connect to others with similar characteristics \cite{firth2020using} \cite{block2020social}.

The general heritable mechanism is as follows: Once a node leaves, it is removed from the network while its connections will be inherited totally or partially by another existing node that is the most similar to it in the network. In other words, a node (inheritor node) in the network will connect to the removed node's original neighbors. The similarity between two nodes can be measured by different certificates such as the Pearson correlation coefficient. Specifically, we proposed three disconnection mechanisms, (i) brutal deletions, (ii) totally heritable deletions, (iii) partially heritable deletions in our primer work. Whereas we focus on the effect of heritable disconnections in an epidemic in this paper, for simplicity, we assume the totally heritable disconnection and randomly select a inheritor node.




The evolution of the birth-death network has a great influence on epidemic spreading on the network, especially the heritable disconnection plays a significant role in epidemic spreading. Based on our proposed evolving network model, we next study the epidemic spreading process on this kind of evolving network.
\subsection{Descriptions of Epidemic Spreading with Indirect Contact}\label{sec:model}
We next focus on the impact of the network evolution our evolving network model on epidemic transmissions and reveal the practical significance of our evolving network model on epidemic spreading among individuals.

\begin{figure}
    \centering
    \begin{minipage}{0.48\textwidth}
        \centering
        \includegraphics[width=0.9\textwidth]{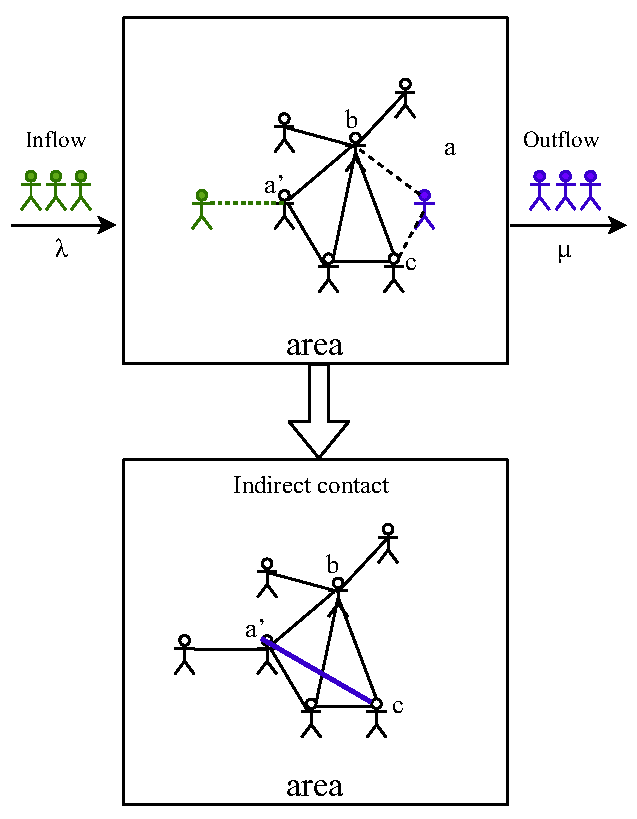}
        \subcaption{}
    \end{minipage}%
    \hfill
    \begin{minipage}{0.48\textwidth}
        \centering
        \includegraphics[width=0.935\textwidth]{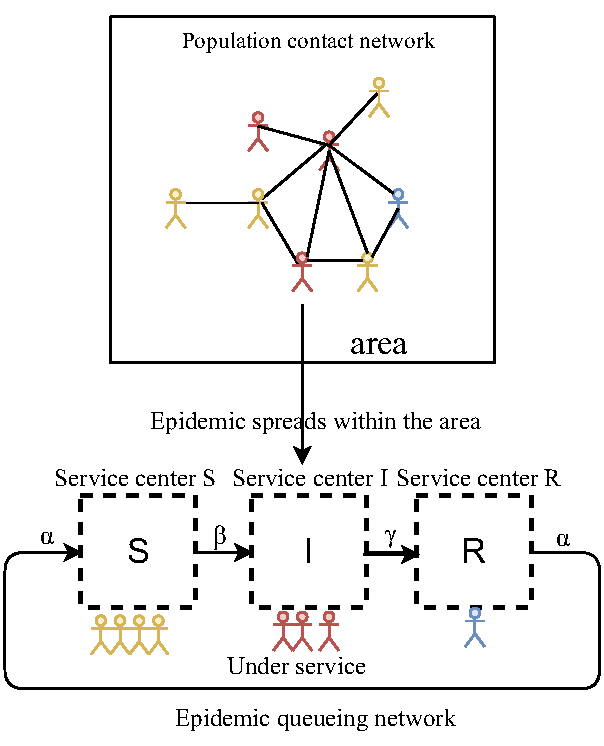}
        \subcaption{}
    \end{minipage}
    \caption{Illustration of the network evolution and epidemic spreading on the evolving network. Sub-figure (a) presents the process of a leaving individual transferring their connections and creating indirect contacts with other individuals. The individual colored green is a newly-coming one who connects to an existing individual a'. The individual colored purple (labeled as a) is the one who are leaving, and dashed lines represent disconnected edges (individual a is disconnecting with b and c). According to the heritable deletion mechanism, the connections of a, i.e., the edge between a and b and the edge between a and c are transferred to a'. As a result, there will be a new edge (contact) between a' and c. Besides, since a' and b are connected originally, there will not be a multi-edge between them. Sub-figure (b) presents an epidemic queueing system, where three epidemic states, i.e. $S$, $I$, and $R$ are three service centers. Susceptible, infected, and recovered individuals are colored yellow, red, and blue.}
    \label{fig:1}
\end{figure}

The evolution of our network is capable of describing the migration of the mobile population. We assume that an evolving network represents a contact population network of an area or a region including mobility. The input of nodes indicates the inflow of individuals into this area, while the output of nodes indicates the outflow of individuals out of this area. For the inflow, an individual coming to this area connects to existing individuals via $m$ edges following the preferential attachment. Once they are connected, the new individual and the existing ones can transform diseases via connections. For the leaving process, an individual leaves the area which will be removed from the network, while their connections with others will be inherited by another existing individual.

For the epidemic spreading process, we utilize the susceptible-infected-recovered-susceptible (SIRS) model which includes the failure of immunity. This suggests three transitions between the three phases, susceptible (S), infected (I), and recovered (R). In detail, we assume that a susceptible individual may be infected when contacting an infected individual, and then becomes infected at an infected rate $\beta$. While, an infected individual can be recovered, obtaining the immunity of the disease at a recovered rate $\gamma$, and can not be infected once becoming recovered until revive to become susceptible again at a reviving rate $\alpha$ because of the immune elimination.

For a better understanding of epidemic spreading on our proposed evolving network, we display it in four parts as follows.

\noindent\textbf{(1) Initialization}

The initial network is a WS small-world network representing the population contact network of an area, which includes one infected individual, while others are all susceptible.

\noindent\textbf{(2) Inflow of individuals}

Individuals come to the area randomly at a rate $\lambda$. Hence the inflow of individuals is a Poisson flow $\Lambda(t+\tau)$ with a parameter $\lambda$, following
\begin{equation}
P\{\Lambda(t+\tau)-\Lambda(t)=n\}=\frac{e^{-\lambda \tau}(\lambda \tau)^{n}}{n!}
\end{equation}
where $P\{\Lambda(t+\tau)-\Lambda(t)=n\}$ is the probability of the number of individuals inflowing during time $\tau$ being $n$.

Once an individual arrives in this area, he connects to $m$ existing individuals by the probability in Eq. \ref{pa}, which indicates that they have contacts via which disease can spread between them.

\noindent\textbf{(3) Transition among epidemic phases}

We consider a Susceptible-Infected-Recovered-Susceptible (SIRS) compartmental model to simulate the spread of epidemic. In detail, if a susceptible individual is connected to an infected one, they get infected at a rate $\beta$. Infected individuals become recovered at a rate $\gamma$, and recovered individuals can become susceptible again at a rate $\alpha$. Notably, $\beta$, $\gamma$, and $\alpha$ are rates rather than probabilities. We choose the SIRS model to simulate the epidemic processes since it has been largely used to model infectious diseases considering a decay of immunity, such as influenza, COVID-19, dengue fever \cite{yuan2021modeling} \cite{pei2017counteracting}
\cite{arsene2022modeling} \cite{aogo2023effects} etc. We assume that there are an input rate and an output rate for each epidemic phase, and the input and output of individuals of an epidemic phase both follow Poisson processes. Individuals go into phase $j$ at the input rate $\lambda_j$ of phase $j$ and leave phase $j$ at the output rate $\mu_j$ of phase $j$. For instance, susceptible individuals becoming infected suggests that they leave the $S$ phase and enter the $I$ phase. Therefore, the output rate of phase $S$ is the input rate of phase $I$, the output rate of phase $I$ is the input rate of phase $R$, and the input rate of phase $S$ is equal to the output rate of phase $R$. This constructs a Markovian queueing network \cite{borovkov1987limit} for the epidemic process, where an epidemic phase is a server, individuals are customers, and being in phase $j$ is regarded as under the service at server $j$. Besides, if an individual is under phase $j$, the time of being in phase $j$ is exponentially distributed with the parameter consistent with the output rate $\mu_j$ of phase $j$. Fig. \ref{fig:1} (b) illustrates the above process.

\noindent\textbf{(4) Outflow of individuals}

The outflow of population occurs randomly at a rate $\mu$, and individuals in any epidemic state can leave the network at any time. Thus the process of individuals leaving the area is a Poisson process with a parameter $\mu$, following
\begin{equation}
P\{U(t+\tau)-U(t)=n\}=\frac{e^{-U \tau}(U \tau)^{n}}{n!}
\end{equation}
which is the probability that the number of individuals out-flowing is $n$ during time $\tau$
Once an individual leave, the structure of the underlying network changes due to the alteration of connections.


According to the heritable disconnection mechanism, indirect contacts form during epidemic spreading as individuals form new contacts with others from the leaving individuals. In other words, we assume that the construction of indirect contacts occurs once an individual leaves. The connections of a leaving individual $i$ will be transformed to a selected individual in this area. Fig. \ref{fig:1} (a) illustrates how a leaving individual creates a contact between other individuals. As is shown in Fig. \ref{fig:1} (a), the leaving individual a transforms its connections to the other individual a'. The edge between individuals a and b and the edge between individuals a and c are removed. Meanwhile, there is an edge forming between a' and c. Since a' and b are connected originally, there will not be loops and there is still one edge between a' and b. In this way, individual a is an intermediary who creates connections (contacts) between another individual and their neighbors.


Above all, we described epidemic spreading on our evolving network, which considers the dynamics of both epidemic propagation and network evolution. We also revealed how the evolution and the heritable disconnection of the network mechanism impact epidemic processes in a way of reality.
\subsection{Analysis of Population Size of Each Epidemic State}
In this sub-section, we perform analyses for the change of the individual number in three epidemic states, denoted by $S(t)$, $I(t)$, and $R(t)$ respectively.

\begin{figure}[t]
\centering
\includegraphics[scale=0.7]{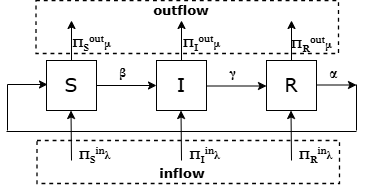}
\caption{Illustration of a queueing network of epidemic states. There are three service centers representing the three epidemic states $S$, $I$, and $R$ marked by square boxes. Arrows between two service centers indicate transitions from one epidemic state to another, and the Greek alphabets ($\beta$, $\gamma$, and $\alpha$) above arrows indicate transition rates between two epidemic states. For each service center, there is an arrow toward it representing the input of individuals of the same epidemic state and an outward arrow representing the output of individuals. The relevant input or output rates are marked on the right of the arrows. The input and output of the three service centers together compose the inflow and outflow of the queueing network.} \label{fig:queueing}
\end{figure}

The queueing network of epidemic states is presented in Fig. \ref{fig:queueing} where the input and output rates of each state are marked. We suppose that the inflow and outflow of individuals include individuals of all three states, and within the inflow, there are more susceptible individuals than infected and recovered ones. Therefore, we set the total individual input rate is $\lambda$, then the inflow rates are respectively $\pi_S^{in}\lambda$, $\pi_I^{in}\lambda$, and $\pi_R^{in}\lambda$ for state $S$, $I$, and $R$, where $\pi_S^{in}$, $\pi_I^{in}$, and $\pi_R^{in}$ are parameters ranging from 0 to 1, and commanding $\pi_S^{in}>>\pi_R^{in}>\pi_I^{in}$. Analogously, the outflow rates of state $S$, $I$, and $R$ are orderly $\pi_S^{out}\mu$, $\pi_I^{out}\mu$, and $\pi_R^{out}\mu$. Besides the inflow and outflow of individuals, there are transitions between states within the queueing network presented in Sec. \ref{sec:model} (3). In the whole queueing network, the transition rate of any one individual from $S$ to $I$ is $I(t)\beta k(t)$, where $ k(t)$ indicates the network average degree at present time $t$, and $\beta$ indicates the property (transmissibility) of the variant, thus it is the same for all individuals. The transition rate of any one individual from $I$ to $R$ is $I(t)\gamma$, and the transition rate of any one individual from $R$ to $S$ is $R(t)\alpha$. Then, the number of individuals in the three epidemic states (customers in three service centers) is respectively a Markov process.
We first deduce the probability that a new individual coming and the probability that an individual leaving of each epidemic state in Lem. \ref{lm:1}

\begin{Lemma}\label{lm:1}
In the SIRS epidemic spreading process on our proposed evolving network, the probability that an individual coming to state $S$, $I$, and $R$, denoted by $P_{*, in}$, is respectively
\begin{equation}
\left\{
\begin{aligned}
P_{S,in}&=[R(t)\alpha+\pi_S^{in}\lambda]\Delta t+o(\Delta t) \\
P_{I,in}&=[S(t)I(t)\beta k(t)+\pi_I^{in}\lambda]\Delta t+o(\Delta t) \\
P_{R,in}&=[I(t)\gamma+\pi_R^{in}\lambda]\Delta t+o(\Delta t) \\
\end{aligned}
\right.
,
\end{equation}
the probability that an individual leaving state $S$, $I$, and $R$, denoted by $P_{*, out}$, is respectively
\begin{equation}
\left\{
\begin{aligned}
P_{S,out}&=S(t)[I(t)\beta k(t)+\pi_{S}^{out}\mu]\Delta t+o(\Delta t) \\
P_{I,out}&=I(t)(\gamma+\pi_{I}^{out}\mu]\Delta t+o(\Delta t) \\
P_{R,out}&=R(t)(\alpha+\pi_{R}^{out}\mu)\Delta t+o(\Delta t) \\
\end{aligned}
\right.
\end{equation}
\end{Lemma}

\begin{Proof}\normalfont
The input of state $S$ includes two parts, the inflow of susceptible individuals from the outside system, whose probability is $\pi_S^{in}\lambda\Delta te^{-\pi_S^{in}\lambda\Delta t}$, and recovered individuals reviving to state $S$, of which the probability is $1-e^{-R(t)\alpha\Delta t}$. In light of the law of total probability, the probability that a new individual comes to state $S$ $P_{S,in}$ is
\begin{equation}\label{eq:Psin}
\begin{aligned}
P_{S,in}&=\pi_S^{in}\lambda\Delta te^{-\pi_S^{in}\lambda\Delta t}\cdot e^{-R(t)\alpha\Delta t}+[1-\pi_S^{in}\lambda\Delta te^{-\pi_S^{in}\lambda\Delta t}][1-e^{-R(t)\alpha\Delta t}]\\
&=[R(t)\alpha+\pi_S^{in}\lambda]\Delta t+o(\Delta t)
\end{aligned}
\end{equation}

The output of the state $S$ also includes the susceptible leaving the system and transforming to state $I$, the probability of which, denoted by $P_{S,out}$ is
\begin{equation}
\begin{aligned}
P_{S,out}&=\sum_{m=1}^{S(t)}(e^{-I(t)\beta k(t)\Delta t}\cdot e^{\pi_S^{out}\mu\Delta t})^{m-1}\cdot (1-e^{-I(t)\beta k(t)\Delta t}\cdot e^{\pi_S^{out}\mu\Delta t})\\
&=S(t)[I(t)\beta k(t)+\pi_{S}^{out}\mu]\Delta t+o(\Delta t)
\end{aligned}
\end{equation}

Analogously, the input of the state $I$ includes susceptible individuals transforming to the state $I$, whose probability is $1-e^{S(t)I(t)\beta k(t)\Delta t}$, as well as the inflow of infected individuals from outside the system during $\Delta t$, whose probability is. The probability that one susceptible individual finishes the service in state $S$ and transforms to state $I$ is $\pi_I^{in}\lambda\Delta te^{-\pi_I^{in}\lambda\Delta t}$. Consequently, the probability that a new individual comes to state $I$ during $\Delta t$, marked as $P_{I,in}$ is
\begin{equation}\label{eq:Piin}
\begin{aligned}
P_{I,in}&=\pi_I^{in}\lambda\Delta te^{-\pi_I^{in}\lambda\Delta t}\cdot e^{S(t)I(t)\beta k(t)\Delta t}+[1-\pi_I^{in}\lambda\Delta te^{-\pi_I^{in}\lambda\Delta t}][1-e^{S(t)I(t)\beta k(t)\Delta t}]\\
&=[S(t)I(t)\beta k(t)+\pi_I^{in}\lambda]\Delta t+o(\Delta t)
\end{aligned}
\end{equation}

The output of the state $I$ includes the outflow of infected individuals out of the system and infected individuals transiting to state $R$. Hence, the probability that one individual leaves state $I$, denoted by $P_{I,out}$, is
\begin{equation}\label{eq:piout}
\begin{aligned}
P_{I,out}&=\sum_{m=1}^{I(t)}(e^{-\gamma\Delta t}\cdot e^{\pi_I^{out}\mu\Delta t})^{m-1}\cdot (1-e^{-\gamma\Delta t}\cdot e^{\pi_I^{out}\mu\Delta t})\\
&=I(t)(\gamma+\pi_{I}^{out}\mu]\Delta t+o(\Delta t)
\end{aligned}
\end{equation}

Similarly, the probability that a new individual comes to state $R$ is
\begin{equation}\label{eq:Prin}
\begin{aligned}
P_{R,in}&=\pi_R^{in}\lambda\Delta te^{-\pi_R^{in}\lambda\Delta t}\cdot e^{I(t)\gamma\Delta t}+[1-\pi_R^{in}\lambda\Delta te^{-\pi_R^{in}\lambda\Delta t}][1-e^{I(t)\gamma\Delta t}]\\
&=[I(t)\gamma+\pi_R^{in}\lambda]\Delta t+o(\Delta t)
\end{aligned}
\end{equation}
The probability that one individual leaves state $R$ is
\begin{equation}\label{eq:prout}
\begin{aligned}
P_{R,out}&=\sum_{m=1}^{R(t)}(e^{-\alpha\Delta t}\cdot e^{\pi_R^{out}\mu\Delta t})^{m-1}\cdot (1-e^{-\alpha\Delta t}\cdot e^{\pi_R^{out}\mu\Delta t})\\
&=R(t)(\alpha+\pi_{R}^{out}\mu)\Delta t+o(\Delta t)
\end{aligned}
\end{equation}
The results follow.
\end{Proof}

Based on Lem. \ref{lm:1}, we next perform analyses for the variation of population sizes of state $S$, $I$, and $R$.
\begin{Theorem}\label{th:1}
In our SIRS epidemic spreading model based on the evolving network, the population size of state $S$, $I$, $R$ varies by the following probabilities respectively:
\begin{equation}
P^{S}_{m,n}(\Delta t)=\left\{
\begin{aligned}
&[R(t)\alpha+\pi_S^{in}\lambda]\Delta t+o(\Delta t), n=m+1 \\
&S(t)[I(t)\beta k(t)+\pi_S^{out}\mu]\Delta t+o(\Delta t), n=m-1 \\
&1-[R(t)\alpha+\pi_S^{in}\lambda+S(t)I(t)\beta k(t)(t)+S(t)\pi_S^{out}\mu]\cdot\Delta t+o(\Delta t), n=m\\
\end{aligned}
\right.
\end{equation}

\begin{equation}
P^{I}_{m,n}(\Delta t)=\left\{
\begin{aligned}
&[S(t)I(t)\beta k(t)+\pi_I^{in}\lambda]\Delta t+o(\Delta t), n=m+1 \\
&I(t)[\gamma+\pi_I^{out}\mu]\Delta t+o(\Delta t), n=m-1 \\
&1-[S(t)I(t)\beta k(t)(t)+\pi_I^{in}\lambda+I(t)\gamma+I(t)\pi_I^{out}\mu]\cdot\Delta t+o(\Delta t), n=m\\
\end{aligned}
\right.
\end{equation}

\begin{equation}
P^{R}_{m,n}(\Delta t)=\left\{
\begin{aligned}
&[I(t)\gamma+\pi_R^{in}\lambda]\Delta t+o(\Delta t), n=m+1 \\
&R(t)(\alpha+\pi_R^{out}\mu)\Delta t+o(\Delta t), n=m-1 \\
&1-[I(t)\gamma+\pi_R^{in}\lambda+R(t)\alpha+R(t)\pi_R^{out}\mu]\cdot\Delta t+o(\Delta t), n=m\\
\end{aligned}
\right.
\end{equation}
where $P^S_{m,n}(\Delta t)$, $P^I_{m,n}(\Delta t)$, $P^R_{m,n}(\Delta t)$ respectively indicates the probability of population size of state $S$, state $I$, and state $R$ changing from $m$ to $n$ during $\Delta t$.
\end{Theorem}

\begin{Proof}\normalfont
The probability that there is no individual finishing the service and leaving state $S$ during $\Delta t$ is $e^{-S(t)(I(t)\beta k(t)+\pi_S^{out}\mu)\Delta t}$. Combined with Eq. \ref{eq:Psin}, the probability that the individuals in State $S$ increases by 1 during $\Delta t$ is
\begin{equation}
\begin{aligned}
P^S_{m,m+1}(\Delta t)&=P_{S,in}\cdot e^{-S(t)[I(t)\beta k(t)(t)+\pi_S^{out}\mu]\Delta t}\\
&=[R(t)\alpha+\pi_S^{in}\lambda]\Delta t+o(\Delta t)
\end{aligned}
\end{equation}

Based on Eq. \ref{eq:Psin}, the probability that the population size in $S$ state decreases by 1 during $\Delta t$ is
\begin{equation}
\begin{aligned}
P^S_{m,m-1}(\Delta t)&=P_{S,out}\cdot \{1-[R(t)\alpha+\pi_S^{in}\lambda]\Delta t+o(\Delta t)\}\\
&=S(t)[I(t)\beta k(t)+\pi_S^{out}\mu]\Delta t+o(\Delta t)
\end{aligned}
\end{equation}

Since the probability of the variation of the population size being over 1 during $\Delta t$ is $o(\Delta t)$, according to the sum of probabilities equal to 1, the probability that the susceptible population size does not change during $\Delta t$ is
\begin{equation}
\begin{aligned}
P^S_{m,m}(\Delta t)&=1-[R(t)\alpha+\pi_S^{in}\lambda+S(t)I(t)\beta k(t)+S(t)\pi_S^{out}\mu]\cdot\Delta t+o(\Delta t)\\
\end{aligned}
\end{equation}

The probability that no individuals leaves state $I$ is $e^{-(I(t)\gamma+\pi_I^{out}\mu)\Delta t}$. In light of Eq. \ref{eq:Piin}, the probability that the population size of state $I$ increases by 1 during $\Delta t$ is
\begin{equation}\label{eq:i+}
\begin{aligned}
P^I_{m,m+1}(\Delta t)&=P_{I,in}\cdot e^{-(I(t)\gamma+\pi_I^{out}\mu)\Delta t}\\
&=[S(t)I(t)\beta k(t)+\pi_I^{in}\lambda]\Delta t+o(\Delta t)
\end{aligned}
\end{equation}

Together with Eqs. \ref{eq:Piin} and \ref{eq:piout}, the probability that the population size of state $I$ decreases by 1 during $\Delta t$ is
\begin{equation}\label{eq:i-}
\begin{aligned}
P^I_{m,m-1}(\Delta t)&=P_{I,out}\cdot \{1-[S(t)I(t)\beta k(t)+\pi_I^{in}\lambda]\Delta t\}\\
&=I(t)[\gamma+\pi_I^{out}\mu]\Delta t+o(\Delta t)
\end{aligned}
\end{equation}
By Eqs. \ref{eq:i+}, \ref{eq:i-} and the sum of probabilities equal to 1, the probability that the population size of state $I$ does not change during $\Delta t$ is
\begin{equation}
\begin{aligned}
P^I_{m,m}(\Delta t)&=1-[S(t)I(t)\beta k(t)+\pi_I^{in}\lambda+I(t)\gamma+I(t)\pi_I^{out}\mu]\cdot\Delta t+o(\Delta t)\\
\end{aligned}
\end{equation}

Analogously, we can obtain the probabilities of the state $R$ population size increasing by 1-$P^R_{m,m+1}$, decreasing by 1-$P^R_{m,m-1}$ as well as keeping remaining $P^R_{m,m}$.

The results follow.
\end{Proof}

The above lemma and theorem show how the number of individuals in states $S$, $I$, and $R$ change with time by probabilities, based on which we can simulate epidemic spreading given initial settings in simulations.

Based on the above dynamics of epidemics and network evolution, we next deduce the reproductive number of the disease spreading on our evolving network.

\begin{Theorem}\label{th:2}
The reproductive number $\mathcal{R}_0$ of the disease spreading on our evolving network in the stationary state with an average degree $k$ is
\begin{equation}
\mathcal{R}_0 = \frac{\lambda\beta k[\pi_R^{in}\alpha + \pi_S^{in}(\alpha + \pi_R^{out}\mu)]}{\pi_S^{out}\mu(\alpha + \pi_R^{out}\mu)(\gamma + \pi_I^{out}\mu)}
\end{equation}
\end{Theorem}

\begin{Proof}\normalfont
We consider the deterministic equations of the system:
\begin{align}
\frac{dS}{dt} &= R(t)\alpha + \pi_S^{in}\lambda - S(t)\pi_S^{out}\mu - S(t)I(t)\beta k \\
\frac{dI}{dt} &= S(t)I(t)\beta k + \pi_I^{in}\lambda - I(t)\gamma - I(t)\pi_I^{out}\mu \\
\frac{dR}{dt} &= I(t)\gamma + \pi_R^{in}\lambda - R(t)\alpha - R(t)\pi_R^{out}\mu\\
\frac{dN}{dt} &= (\pi_S^{in} + \pi_I^{in} + \pi_R^{in})\lambda - (S(t)\pi_S^{out} + I(t)\pi_I^{out} + R(t)\pi_R^{out})\mu
\end{align}
Following the next generation matrix method \cite{van2017reproduction}, we consider only $\frac{dI}{dt}$, and split $\frac{dI}{dt}$ into two terms $F(S,I,R)$ and $V(S,I,R)$ as 
\begin{equation}
F(S, I, R) = S(t)I(t)\beta k(t)
\end{equation}
and
\begin{equation}
V(S, I, R) = I(t)\gamma + I(t)\pi_I^{out}\mu - \pi_I^{in}\lambda
\end{equation}
We consider the disease free equilibrium (DFE) before a disease is introduced into the system, and denote $S*$, $I*$, $R*$, and $N*$ as the steady state. Under DFE ($\pi_I^{in}=0$), we have $I^* = 0$, $N^*=S^*+R^*$. Let $\frac{dR}{dt} = 0$, we get $\pi_R^{in}\lambda - R^*\alpha - R^*\pi_R^{out}\mu = 0$. Thus, 
$R^* = \frac{\pi_R^{in}\lambda}{\alpha + \pi_R^{out}\mu}$.
Let $\frac{dS}{dt} = 0$, we obtain
\begin{equation}
S^* = \frac{R^*\alpha + \pi_S^{in}\lambda}{\pi_S^{out}\mu} = \frac{\pi_R^{in}\lambda\alpha + \pi_S^{in}\lambda(\alpha + \pi_R^{out}\mu)}{\pi_S^{out}\mu(\alpha + \pi_R^{out}\mu)}
\label{eq:S*}
\end{equation}
Consider the DFE, we define
\begin{equation}
F =\frac{\partial F(S,I,R)}{\partial I} = S^* \beta k
\end{equation}
and
\begin{equation}
V = \frac{\partial V(S,I,R)}{\partial I} = \gamma + \pi_I^{out}\mu
\end{equation}
The reproductive number $\mathcal{R}_0$ is defined as $\frac{F}{V}$, thus
\begin{equation}
\mathcal{R}_0 = \frac{S^* \beta k}{\gamma + \pi_I^{out}\mu}
\label{eq:reproductive_number}
\end{equation}
Substitute Eq. \ref{eq:S*} into Eq. \ref{eq:reproductive_number},
we obtain
\begin{equation}
\mathcal{R}_0 = \frac{\lambda\beta k[\pi_R^{in}\alpha + \pi_S^{in}(\alpha + \pi_R^{out}\mu)]}{\pi_S^{out}\mu(\alpha + \pi_R^{out}\mu)(\gamma + \pi_I^{out}\mu)}
\end{equation}
The results follow.
\end{Proof}

\begin{figure*}[htbp]
\centering
\includegraphics[scale=0.07]{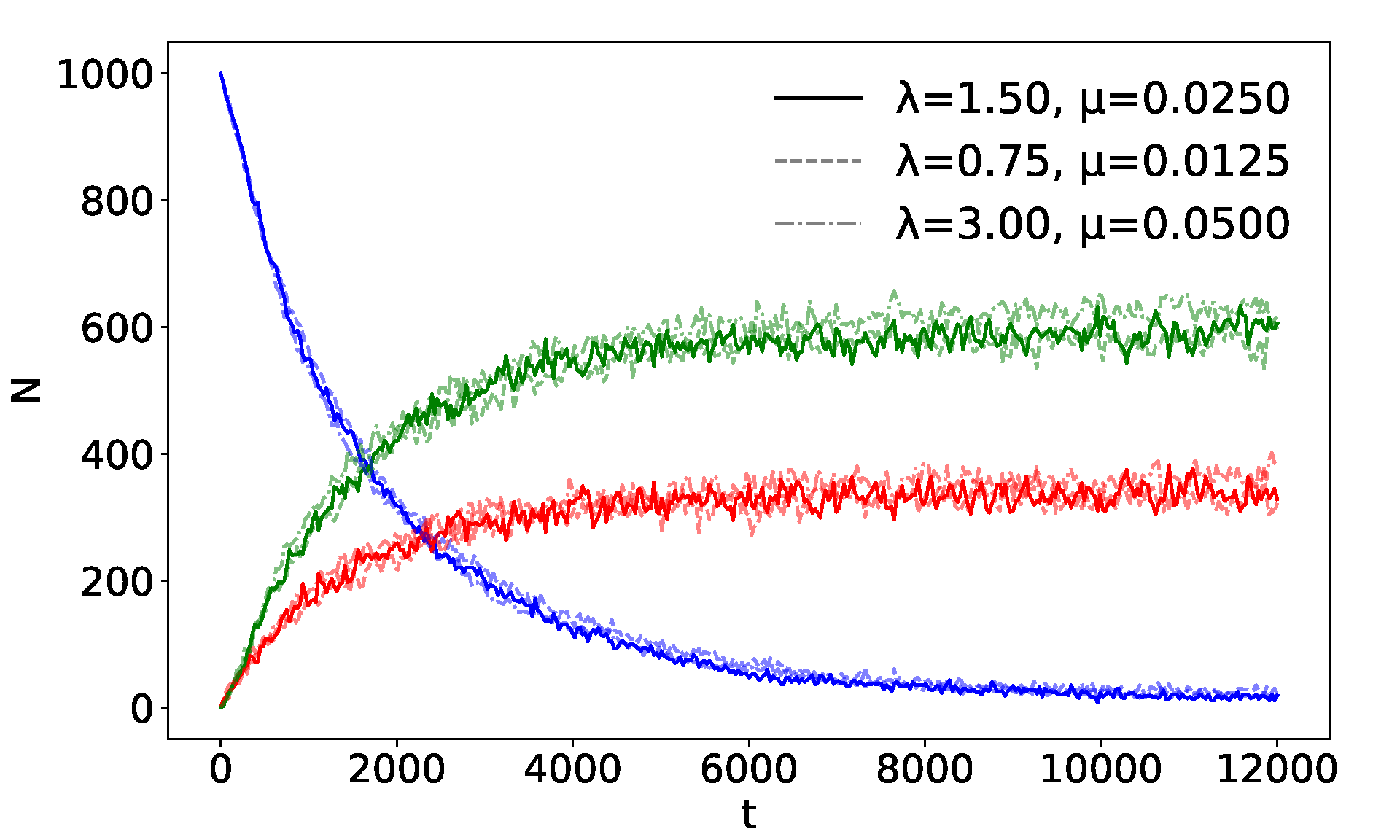}\hspace{0.4cm}
\includegraphics[scale=0.07]{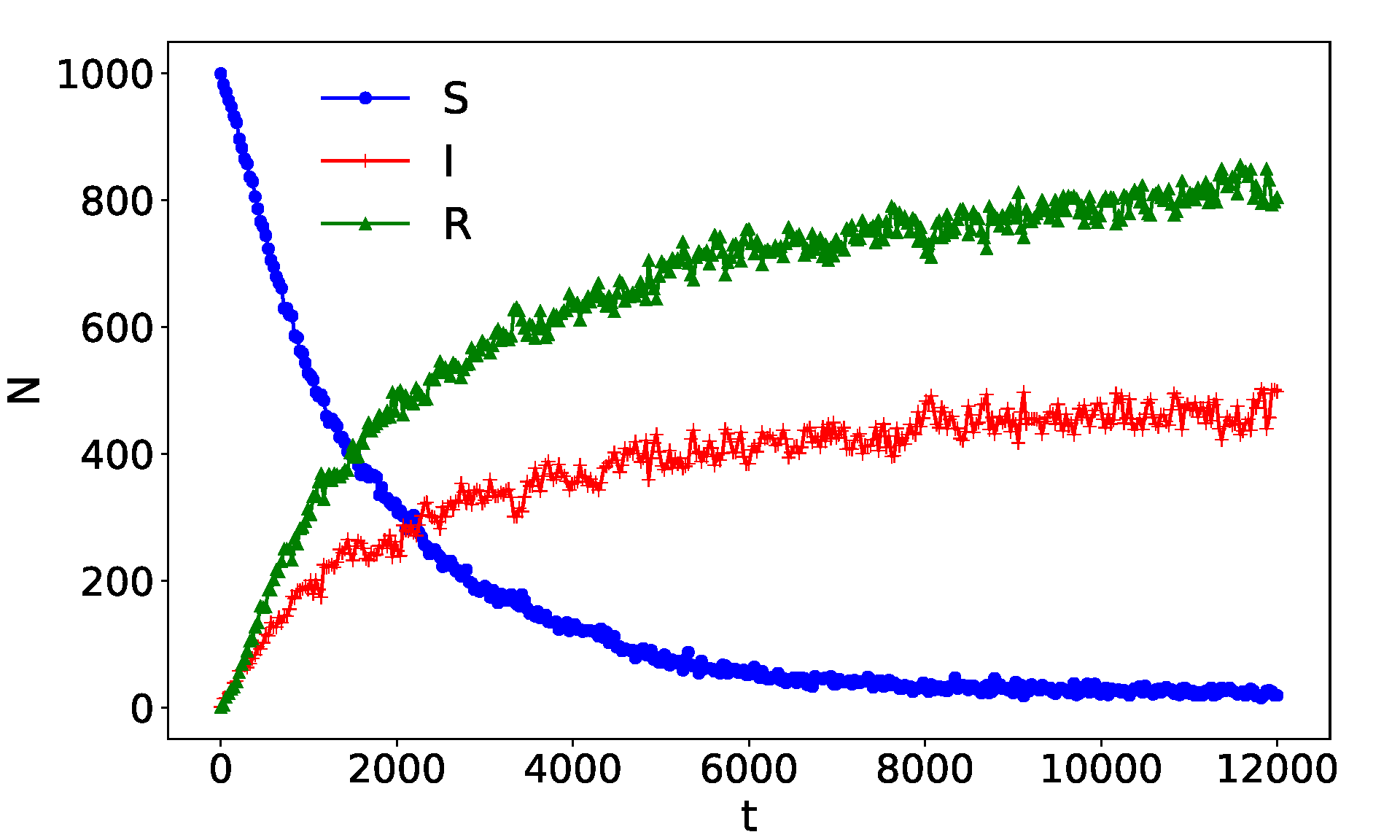}\hspace{0.4cm}
\includegraphics[scale=0.07]{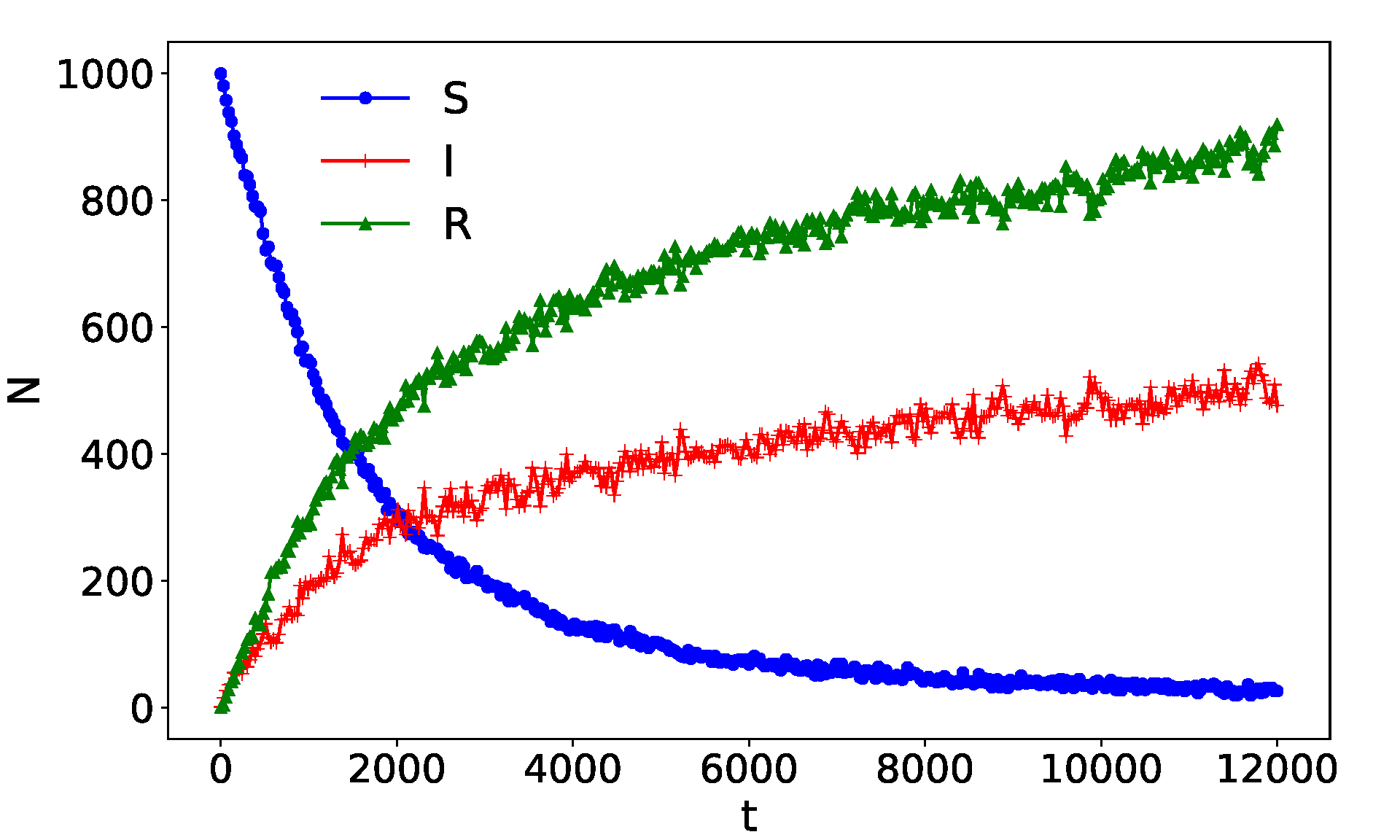}

\begin{tabular}{ccc}
\makebox[4cm]{\footnotesize (a) Three pairs of ($\lambda$, $\mu$)} &
\makebox[4cm]{\footnotesize (b) $\lambda=1.5$, $\mu=0.01$} &
\makebox[4cm]{\footnotesize (c) $\lambda=1.5$, $\mu=0.005$} \\
\end{tabular}

\vspace{0.5cm}

\includegraphics[scale=0.07]{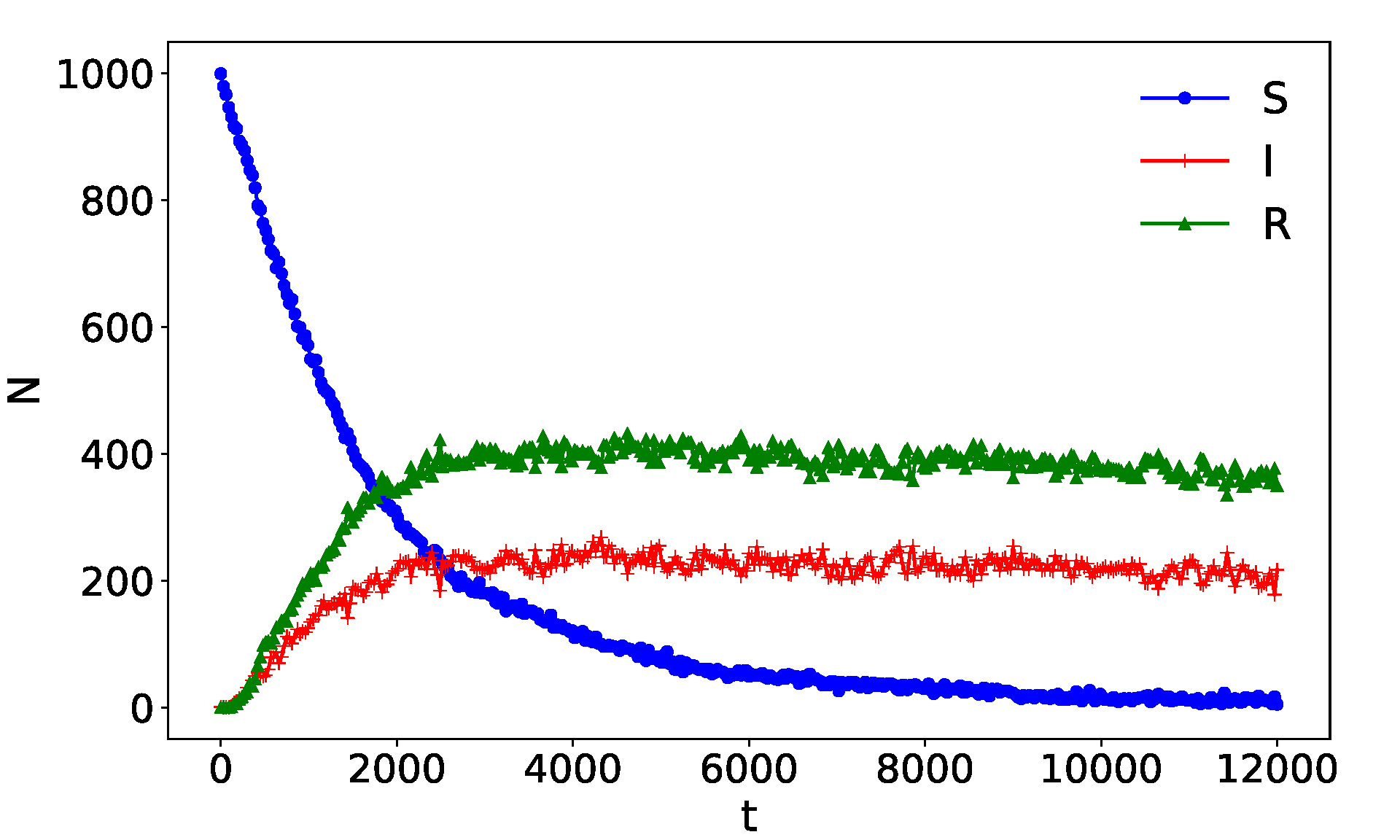}\hspace{0.5cm}
\includegraphics[scale=0.07]{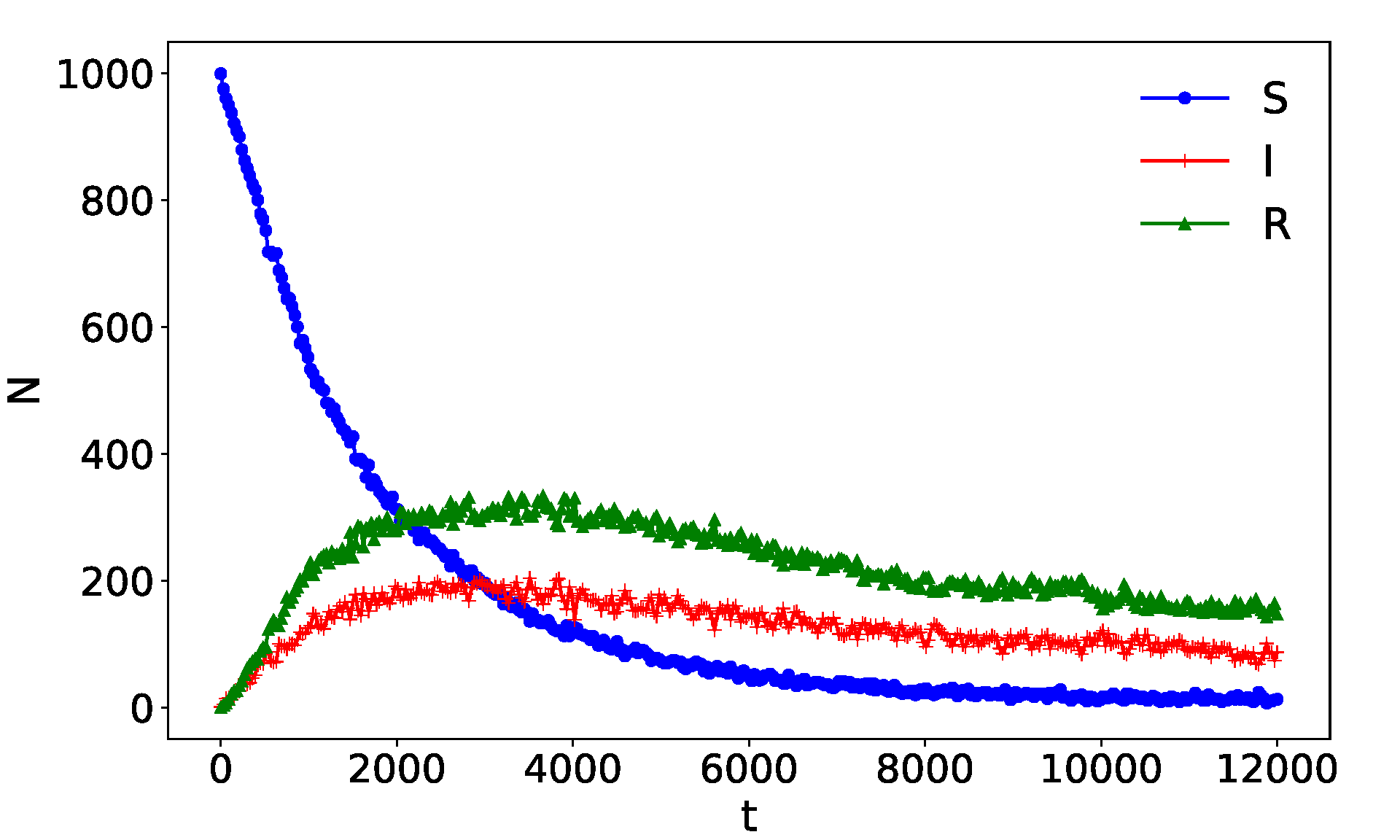}

\begin{tabular}{cc}
\makebox[4cm]{\footnotesize (d) $\lambda=1.5$, $\mu=0.05$} &
\makebox[4cm]{\footnotesize (e) $\lambda=1.5$, $\mu=0.1$} \\
\end{tabular}

\caption{\label{fig:lambmiu}The population size varying with time in state $S$, $I$, and $R$ with different values of the inflow rate $\lambda$ and outflow rate $\mu$. In sub-figure (a), we set three pairs different values of the inflow $\lambda$ and outflow rate $\mu$, (3, 0.05), (1.5, 0.025), (0.75, 0.0125). In these settings, the population size of each state fluctuates horizontally. (b) and (c) present a rising tendency in the population size of state $I$ and $R$ with $\lambda=1.5$, $\mu$ is respectively 0.01 and 0.005. (d) and (e) illustrates a decreasing population size of the whole system with $\lambda=1.5$, $\mu$ is 0.1 and 0.15 subsequently.}
\end{figure*}

\section{Simulation}\label{sec:IV}

\begin{figure*}[t]
\begin{minipage}[t]{0.32\linewidth}
\centering
 \includegraphics[height=3.55cm, width=5.6cm]{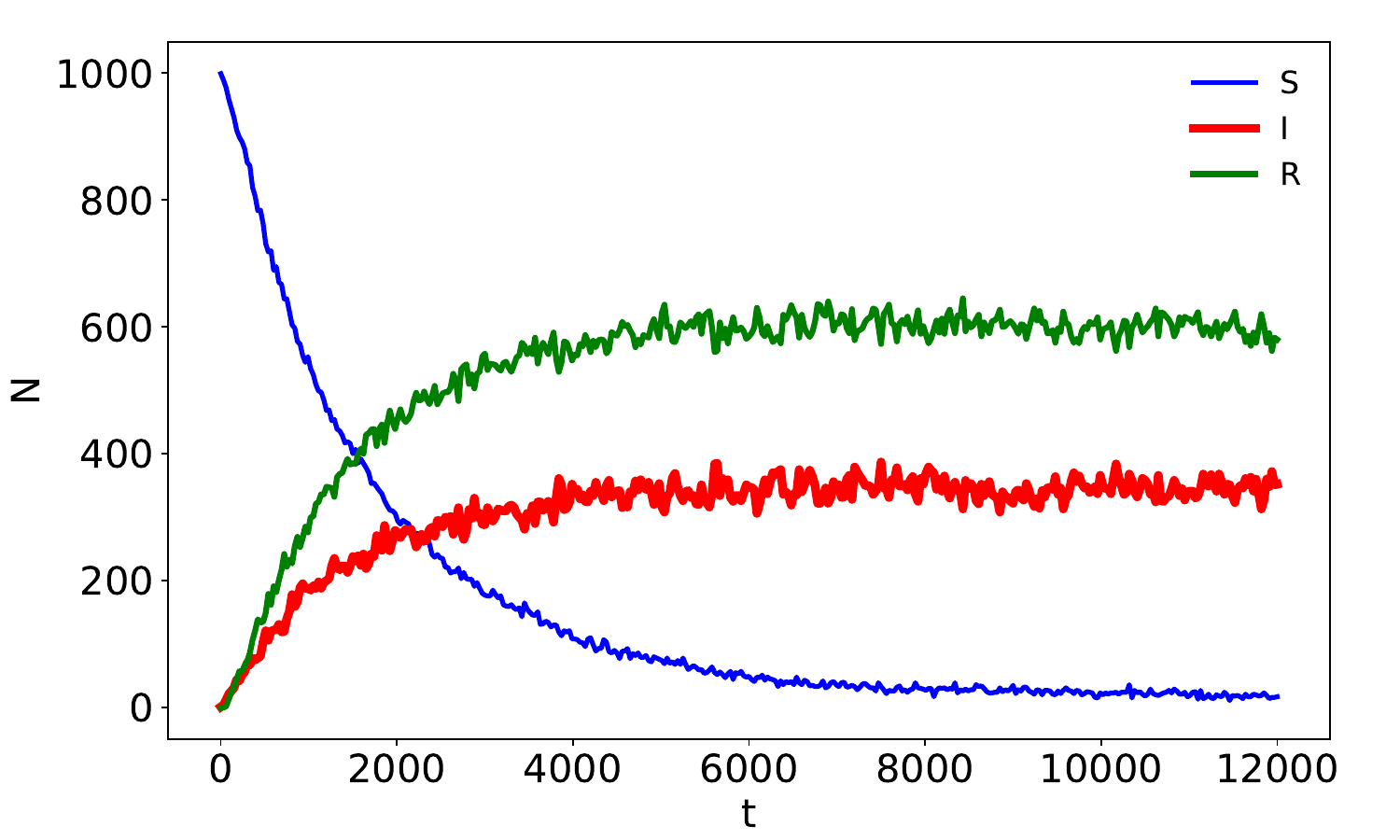}
\parbox{4.5cm}{\centering\footnotesize \hspace{0.1cm}(a) Initial parameter setting}

\end{minipage}
\begin{minipage}[t]{0.32\linewidth}
\centering
 \includegraphics[height=3.55cm, width=5.6cm]{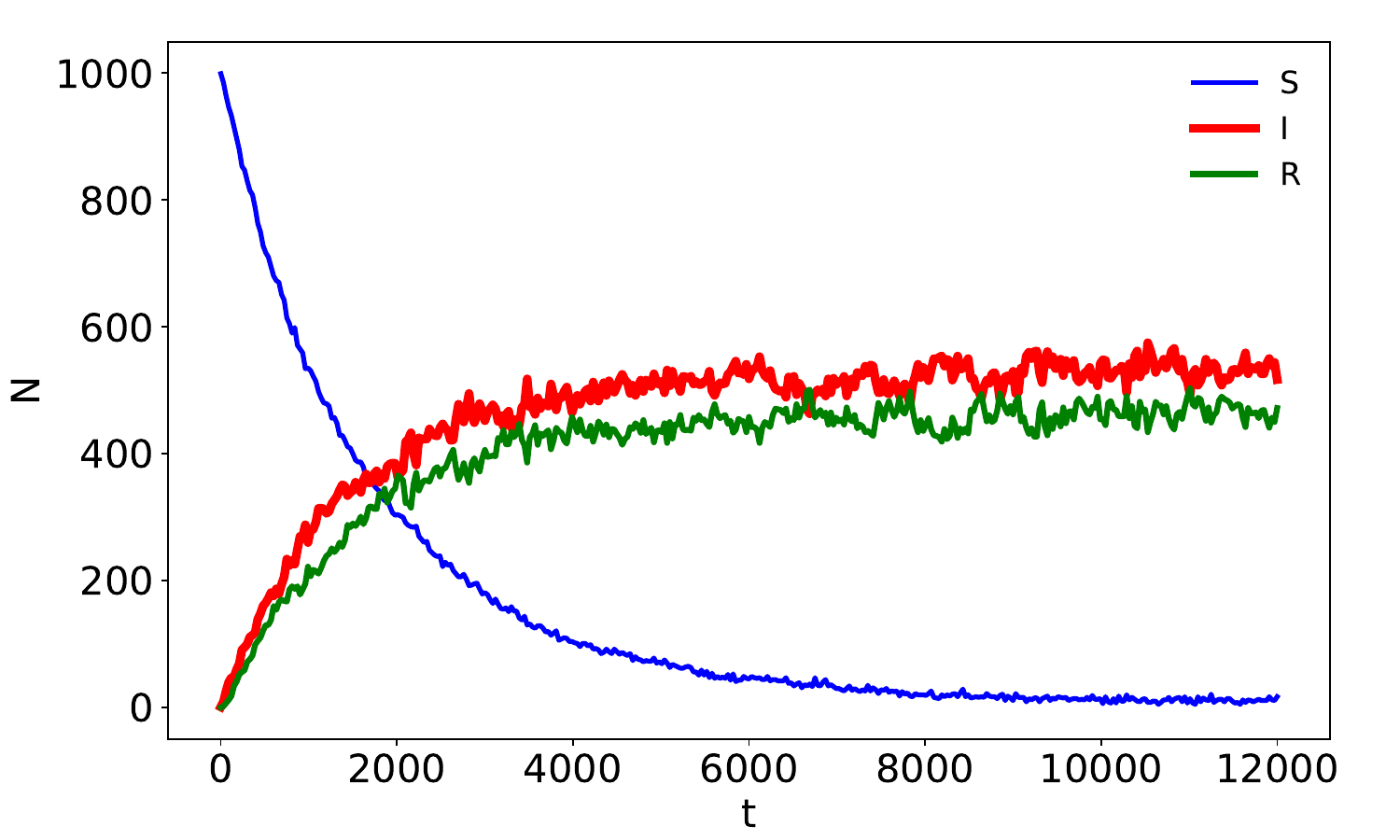}
\parbox{4.5cm}{\centering\footnotesize \hspace{0.1cm}(b) $\gamma$ halved to 0.35}

\end{minipage}
\begin{minipage}[t]{0.32\linewidth}
\centering
 \includegraphics[height=3.55cm, width=5.6cm]{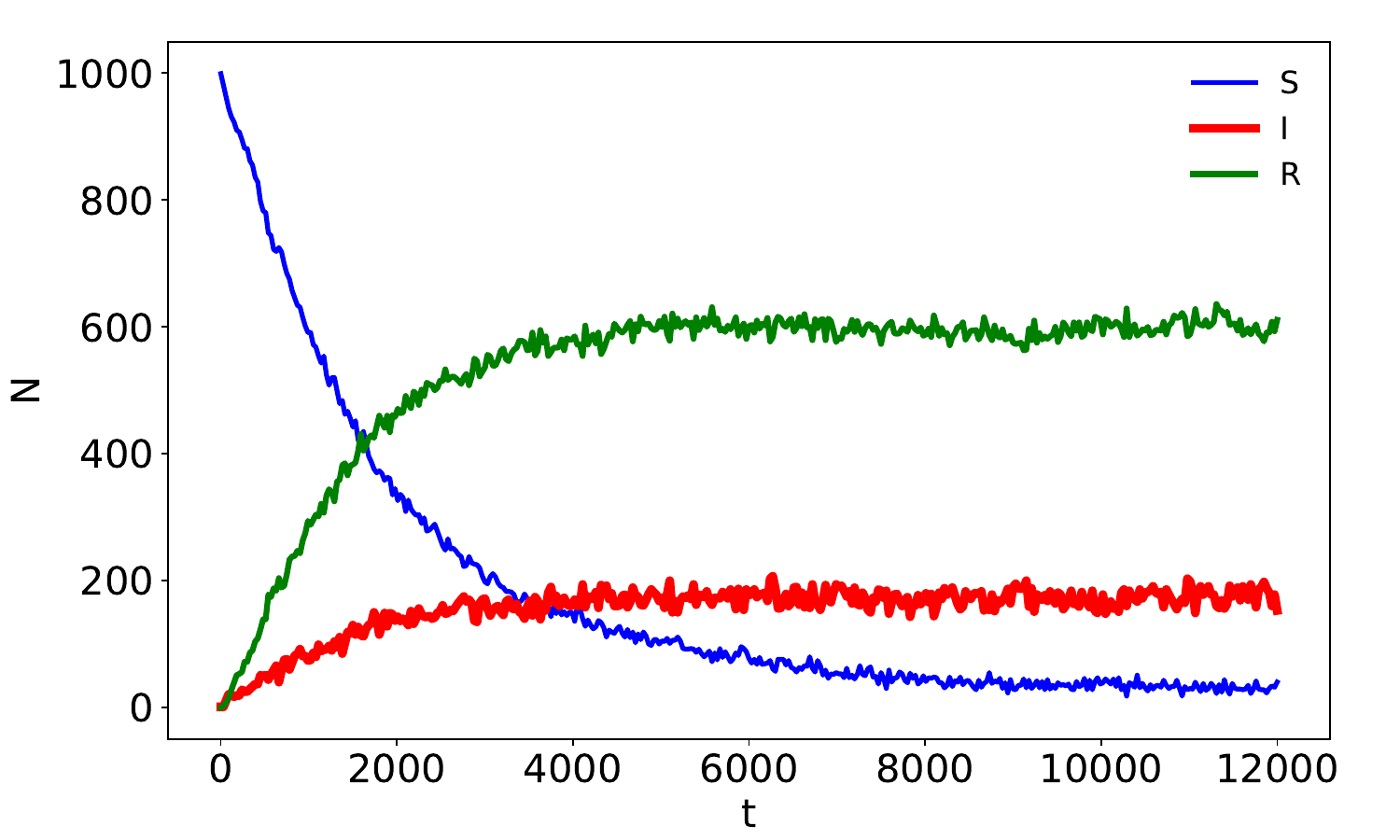}
\parbox{4.5cm}{\centering\footnotesize \hspace{0.1cm}(c) $\gamma$ doubled to 1.4}

\end{minipage}
\begin{minipage}[t]{0.32\linewidth}
\centering
 \includegraphics[height=3.55cm, width=5.6cm]{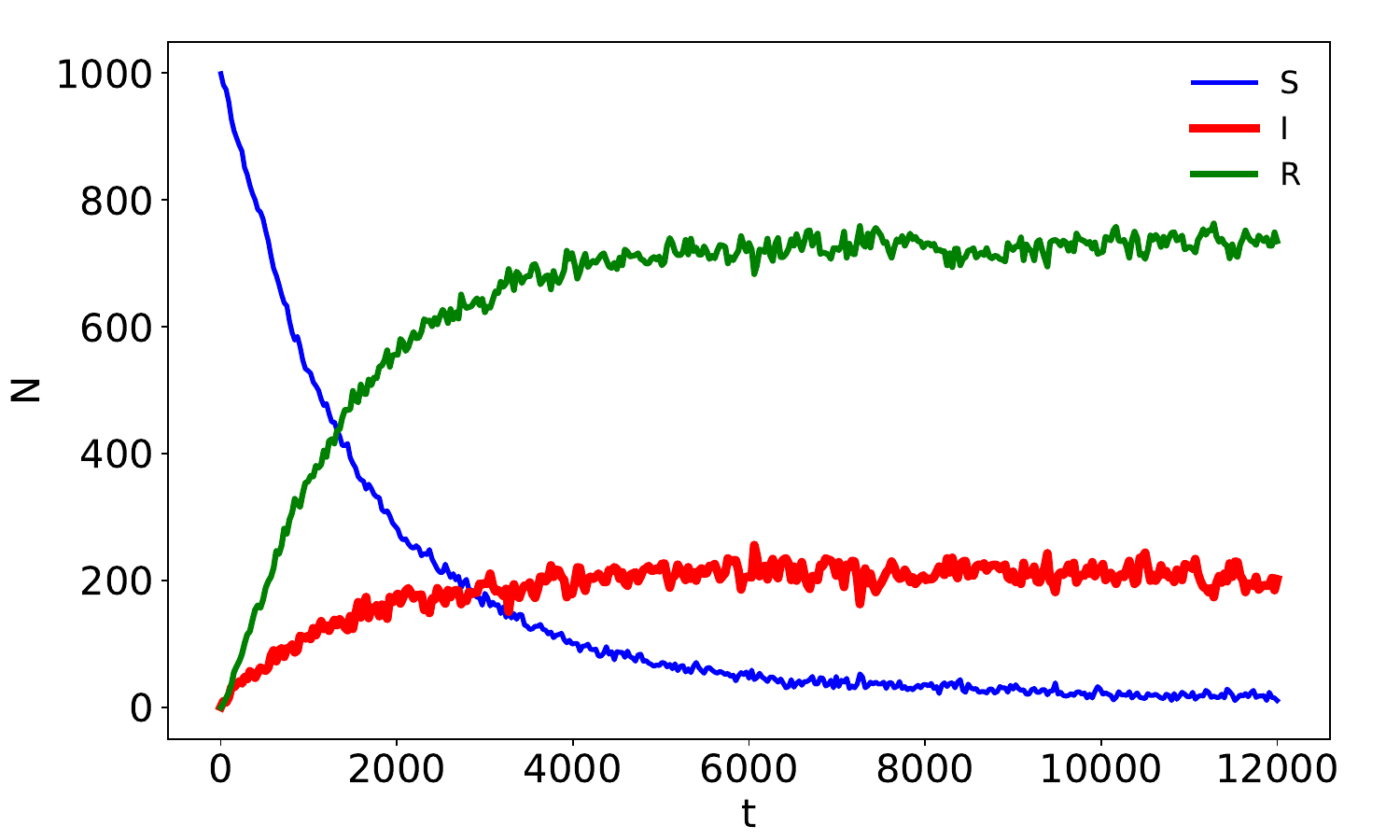}
\parbox{4.5cm}{\centering\footnotesize \hspace{0.1cm}(d) $\alpha$ halved to 0.2 }

\end{minipage}
\begin{minipage}[t]{0.32\linewidth}
\centering
 \includegraphics[height=3.55cm, width=5.6cm]{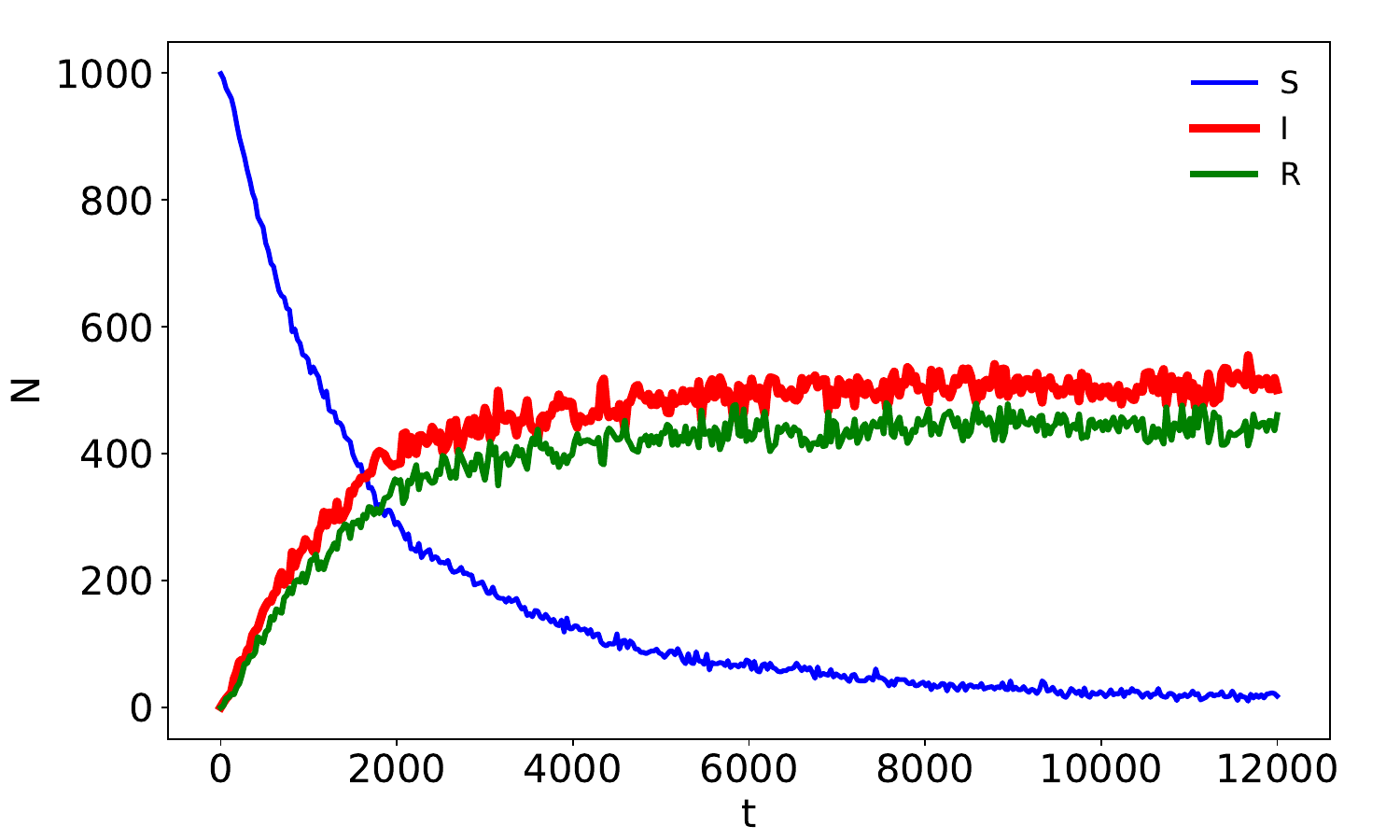}
\parbox{4.5cm}{\centering\footnotesize \hspace{0.1cm}(e) $\alpha$ doubled to 0.8}

\end{minipage}
\begin{minipage}[t]{0.32\linewidth}
\centering
 \includegraphics[height=3.55cm, width=5.6cm]{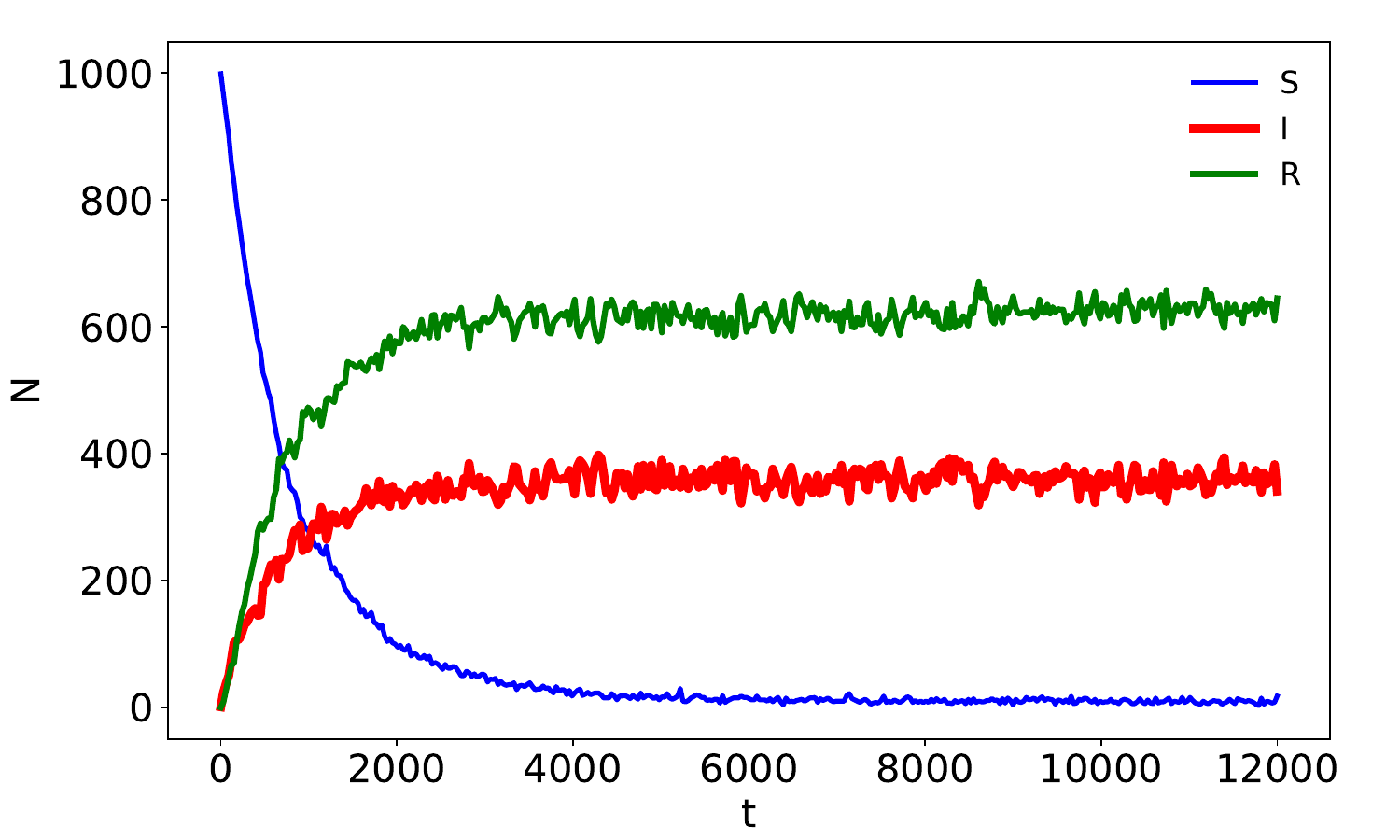}
\parbox{4.5cm}{\centering\footnotesize \hspace{0.1cm}(f) $\beta$ doubled to 0.01}

\end{minipage}
\begin{minipage}[t]{0.32\linewidth}
\centering
 \includegraphics[height=3.55cm, width=5.6cm]{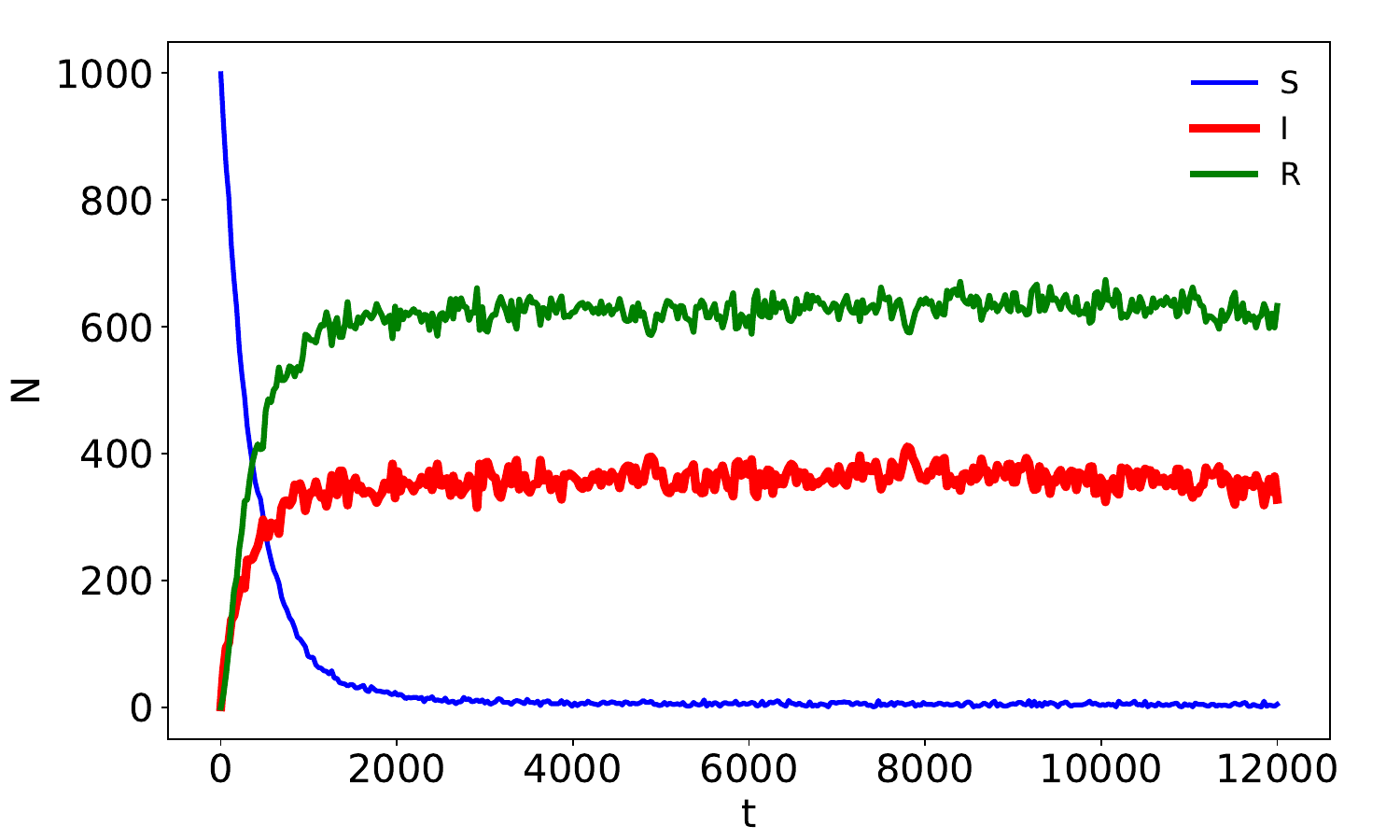}
\parbox{4.5cm}{\centering\footnotesize \hspace{0.1cm}(g) $\beta$ quadrupled to 0.02}
\end{minipage}
\begin{minipage}[t]{0.32\linewidth}
\centering
 \includegraphics[height=3.55cm, width=5.6cm]{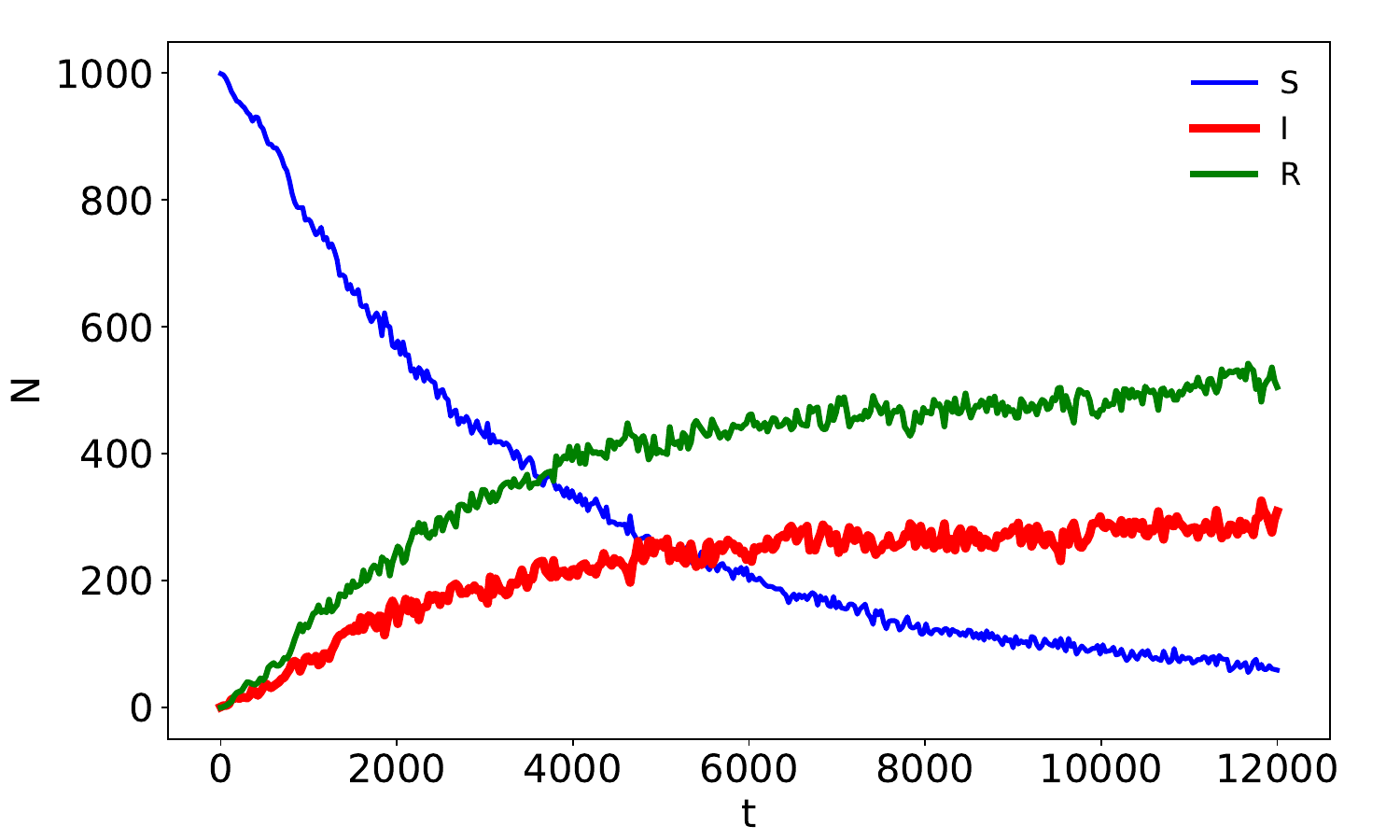}
\parbox{4.5cm}{\centering\footnotesize \hspace{0.1cm}(h) $\beta$ halved to 0.0025}
\end{minipage}
\begin{minipage}[t]{0.32\linewidth}
\centering
 \includegraphics[height=3.55cm, width=5.6cm]{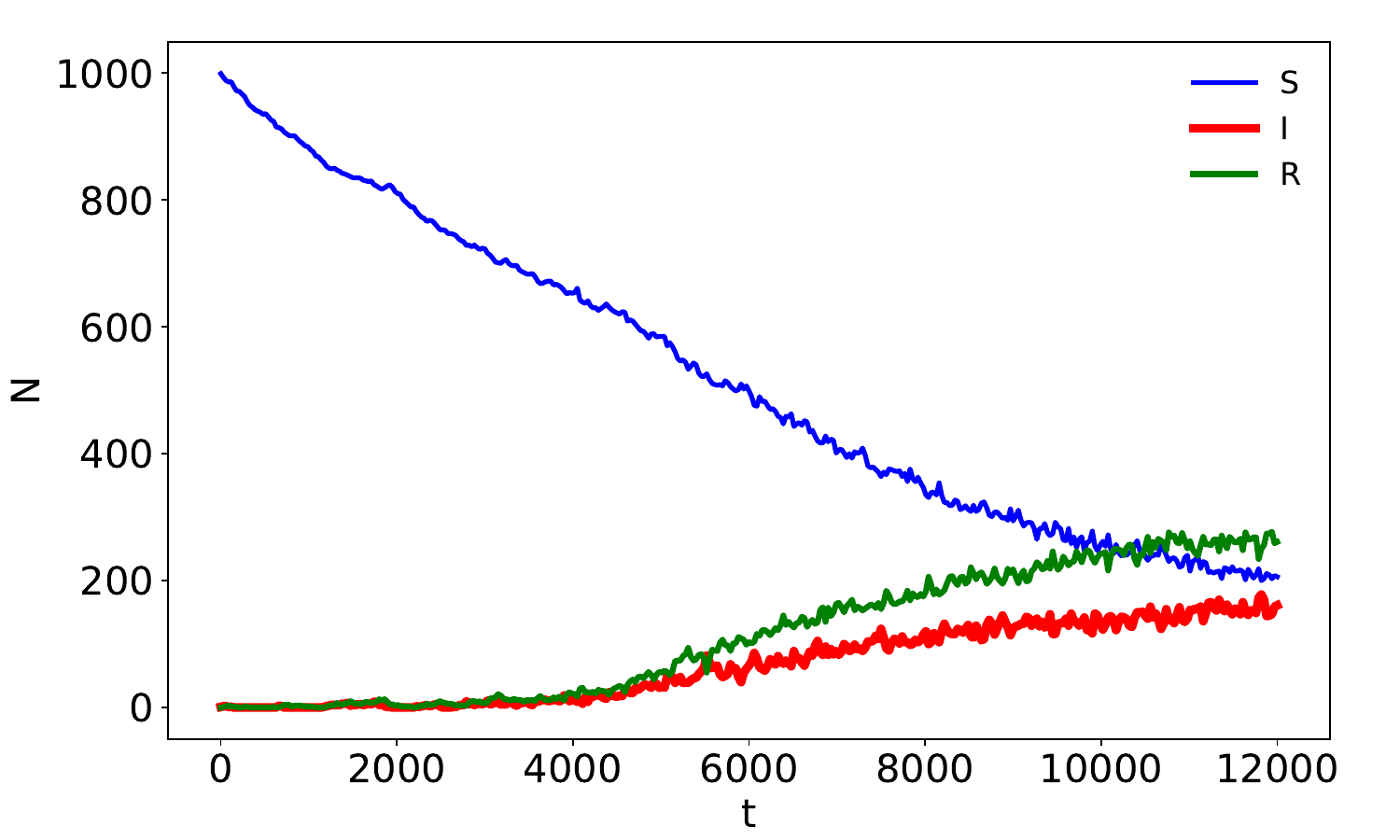}
\parbox{4.5cm}{\centering\footnotesize \hspace{0.1cm}(i) $\beta$ reduced to 0.001}
\end{minipage}
\caption{\label{fig:n123} The number of individuals in the three epidemic compartments $S$, $I$, and $R$ varying with time with different values of the infected rate $\beta$, the recovered rate $\gamma$, and the reviving rate $\alpha$. $\lambda$ and $\mu$ are set as 1.5 and 0.025 in each sub-figure. The initial setting in (a) is $\beta=0.005$, $\gamma=0.7$, $\alpha=0.4$. $\gamma$ is halved in (b) and doubled in (c), and $\alpha$ is halved in (d) and doubled in (e). $\beta$ is heightened to 0.01 and 0.02 respectively in (f) and (g) and reduced to 0.0025 and 0.001 respectively in (h) and (i). Higher $\gamma$ leads to fewer infected and more recovered individuals at steady state. $\gamma$ and $\alpha$ show opposite effects on $I$ and $R$ populations (b-e). Increasing $\beta$ accelerates the approach to steady state without changing final population sizes (f,g), while decreasing $\beta$ delays steady state and reduces equilibrium populations in states $I$ and $R$ (h,i).}
\end{figure*}

In this section, we will display various simulations of epidemic spreading on our evolving networks. we first investigate the impact of the input and output rate on the population size of each epidemic state. Then, we fix the input and output rate and investigate impact of the other epidemic parameters. We further analyze the influence of the heritable mechanism on the epidemic size.
All the simulations are conducted via PYTHON, and we utilize the $networkx$ package to construct networks.

\subsection{Population Size of the Three Epidemic States}\label{sec:A}
The time sequences of the population size in different epidemic states are the most concerned in an epidemic, especially the infected individual number. We first display the population sizes of three epidemic states varying with time under different parameters in our model (SIRS compartmentalization). We conduct numerical experiments according to the rule in Thm. \ref{th:1} and record the number of individuals at each time stamp. For the five parameters, the inflow rate, the outflow rate, the infected rate, the recovered rate, and the reviving rate, we adjust their values and analyze the influence of the different parameters on the population size in the three epidemic states.

The initial network is a WS small-world network with the network size $n$=1000 and the reconnection probability $p$=$0.3$. Each new individual comes with $m$=4 edges which connect to existing ones randomly selected by $choice(list(graph.nodes))$ in PYTHON. The proportions of susceptible, infected, and recovered population within the inflow are set as $\pi_S^{in}$=0.90, $\pi_I^{in}$=0.05, $\pi_R^{in}$=0.05, and the proportions within the outflow are set as $\pi_S^{out}$=0.85, $\pi_I^{out}$=0.05, $\pi_R^{in}$=0.10.

We first adjust the inflow and outflow rate to observe the population size time sequences demonstrated in Fig. \ref{fig:lambmiu}. Apart from $\lambda$ and $\mu$, the parameter setting for the other rates is $\beta$=0.005, $\gamma$=0.7, and $\alpha$=0.4. Fig. \ref{fig:lambmiu} (a) shows the population size under three pairs of ($\lambda$, $\mu$), which are (3, 0.05), (1.5, 0.025), and (0.75, 0.0125) marked by different type lines. The three lines are overlapped most time, which indicates that with the same ratio of $\lambda/\mu$ the population size of each state changes with time by the same discipline, and the population size will be stationary given about $\lambda/\mu$=60. In detail, the population size of state $S$ first decreases while that of state $I$ and $R$ both increase. The three states' population size all keeps stationary after $t$=8000, $S(t)$ fluctuating a little above 0, $I(t)$ and $R(t)$ fluctuating at 300 and 600 approximately. In Figs. \ref{fig:lambmiu} (b) and (c), the infected and recovered population sizes grow with time when lowering the value of $\mu$ compared to sub-figure \ref{fig:lambmiu} (a). Figs. \ref{fig:lambmiu} (d) and (e) present the situation where we raise the value of $\mu$ to 0.1 and 0.15 respectively compared to (a). The population size of both states $I$ and $R$ increases in a short time and then begins to decline slowly in the rest of the time.

Under the stationary situation where the inflow rate $\lambda$=1.5 and $\mu$=0.025, we next adjust the epidemic parameters $\beta$, $\gamma$, and $\alpha$ and analyze their influence on the population sizes of three states, shown in Fig. \ref{fig:n123}. The initial parameter setting is $\beta$=0.005, $\gamma$=0.7, and $\alpha$=0.4, with which the population sizes are plotted in Fig. \ref{fig:n123} (a). In Fig. \ref{fig:n123} (b), the stationary population size of state $I$ increases from about 300 to 500 while the stationary population size of state $R$ decreases from below 600 to 450 approximately when the recovered rate $\gamma$ is reduced to half of the original value. The opposite results are displayed in Fig. \ref{fig:n123} (c) where the stationary population size of $I$ declines from 300 to 200 while the stationary population size of $R$ increases a little, as $\gamma$ doubles. The reason is that $\gamma$ is positively correlated with the output rate of $I$ and the input rate of $S$. In Fig. \ref{fig:n123} (d), the reviving rate $\alpha$ is reduced, and the population size of $I$ decreases from 300 to 200, the population size of $R$ increases by 100 approximately, which is the same as Fig. \ref{fig:n123} (c). Analogously, the plot in Fig. \ref{fig:n123} (e) is the same as that in Fig. \ref{fig:n123} (b). From Figs. \ref{fig:n123} (b)-(e), we see that $\gamma$ and $\alpha$ have the opposite effect on the population size of state $I$ and $R$.

In Figs. \ref{fig:n123} (f) and (g), where we raise the infected rate $\beta$, the population size of state $I$ and $R$ seems the same as that with the initial parameters in Fig. \ref{fig:n123} (a). However, the infected and recovered population grow faster, and the susceptible population declines faster in Figs. \ref{fig:n123} (f) and (g) compared to that in Fig. \ref{fig:n123}. This indicates that the increase of the infected rate $\beta$ advances the time when the population sizes keep stationary advances, from $t$=5000 (in Fig. \ref{fig:n123} (a)) to $t$=3000 with $\beta$=0.01 and to $t$=2000 with $\beta$=0.02. In Figs. \ref{fig:n123} (h) and (i), we lower the value of $\beta$, and the population size of state $I$ and $S$ both decrease. We can also see that the smaller the value of $\beta$ is, the smaller the population size of $I$ and $R$ becomes. This is because a larger infected rate $\beta$ leads to a larger infected and recovered population. Besides, the time when the population size is stationary procrastinates to $t$=7000 with $\beta$=0.0025 in Fig. \ref{fig:n123} (h) and to $t$=30000 with $\beta$=0.001 in Fig. \ref{fig:n123} (g), indicating that the decrease of the infected rate $\beta$ defers the point when the population sizes are stationary. In light of our numerical simulation, $\beta$ affects not only the population sizes but also the stationary time point. Based on the results shown in Figs. \ref{fig:n123} (f)-(i), we find that the influence of the infected rate $\beta$ on the epidemic spreading process is more complex than other parameters. For a better understanding of how the value of $\beta$ affects the spreading process and to investigate the threshold for $\beta$, we next assign $\beta$ to small values even less than the smallest value (0.001) in Fig. \ref{fig:n123} (i), and set the termination time $t$=$1.2\times 10^5$ and $t$=$2.4\times 10^5$ to observe the population size within a quite long time.

\begin{figure*}[htbp]
\begin{minipage}[!t]{0.32\linewidth}
\centering
 \includegraphics[scale=0.08]{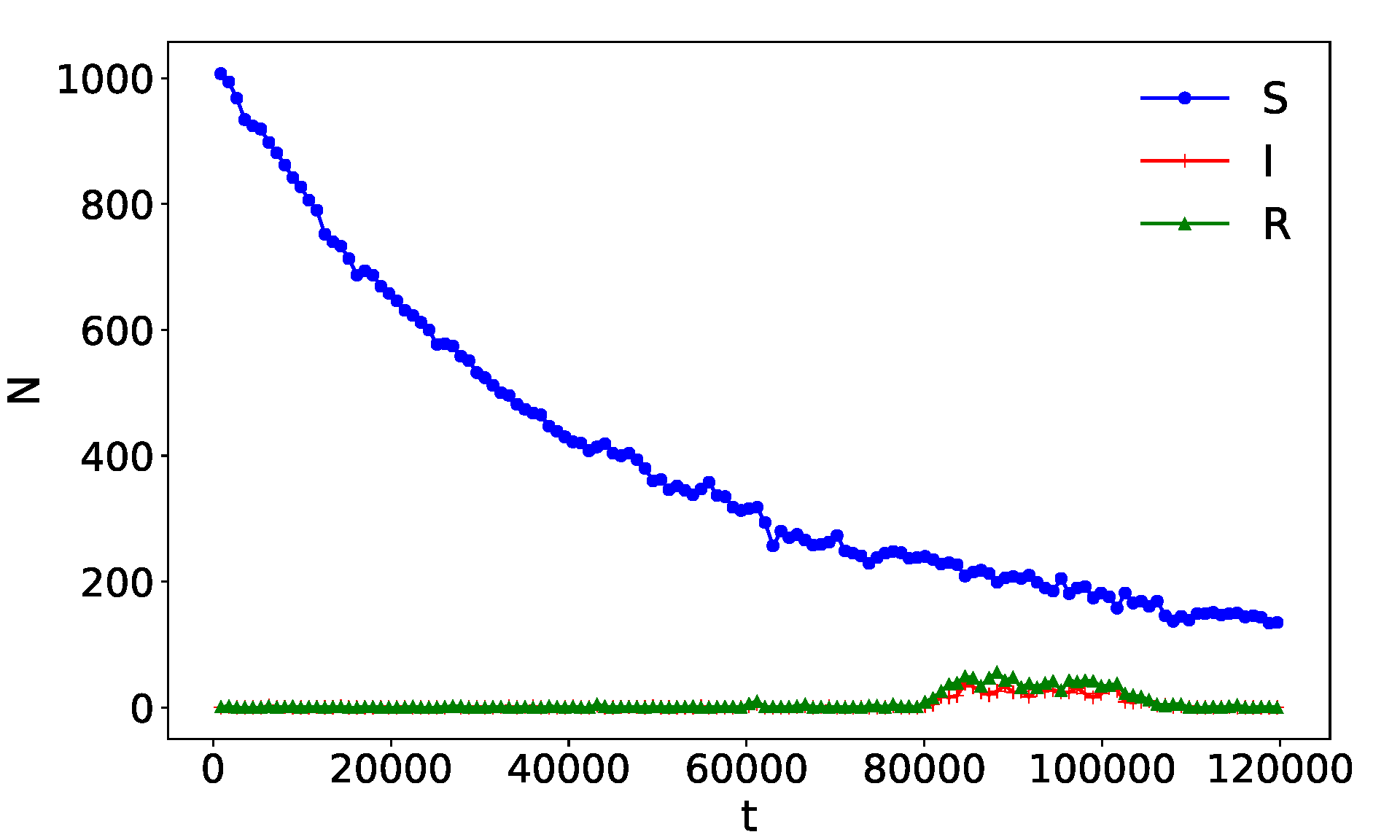}
\parbox{7cm}\centering{\footnotesize (a) $\beta=2\times 10^{-4}$}
\end{minipage}
\begin{minipage}[!t]{0.32\linewidth}
\centering
 \includegraphics[scale=0.08]{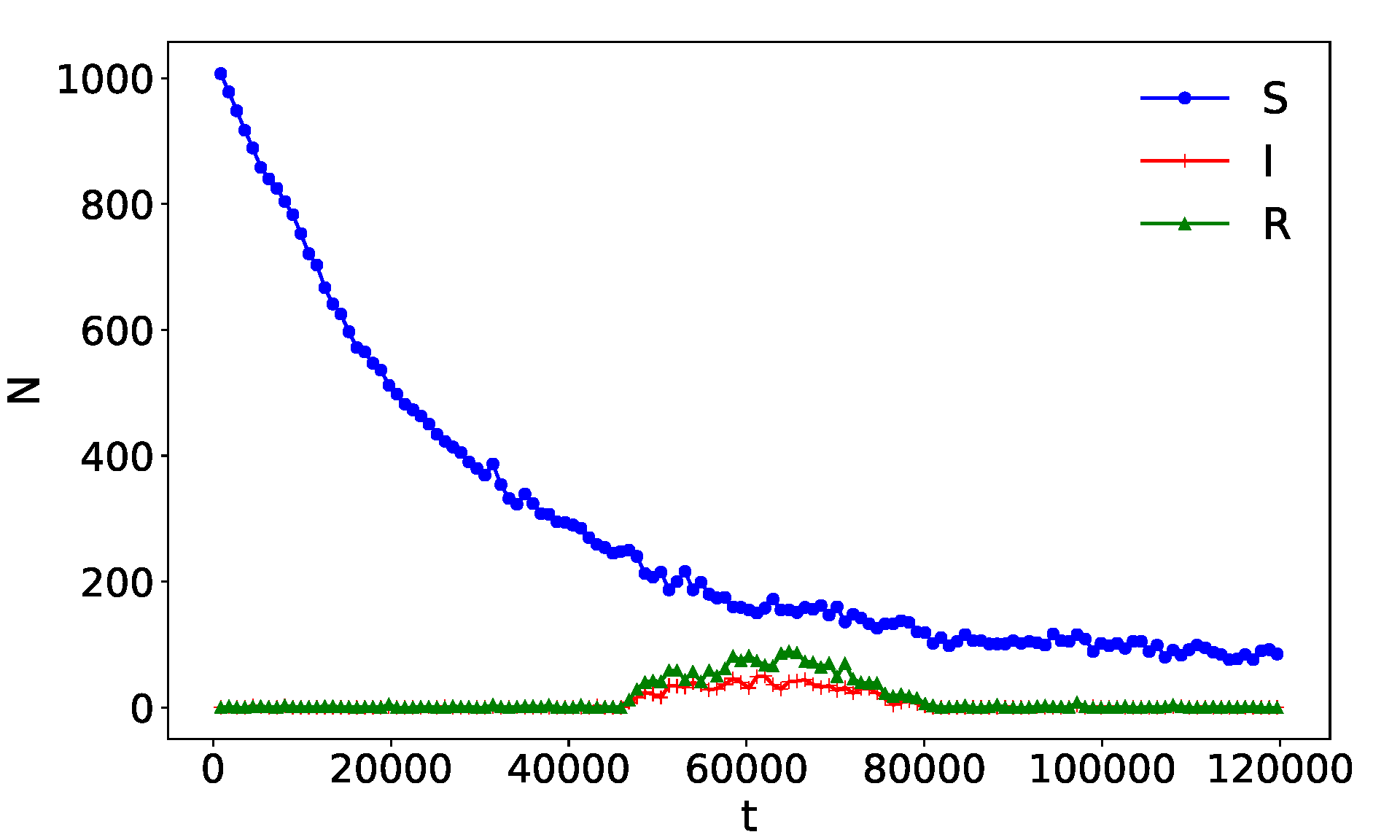}
\parbox{7cm}\centering{\footnotesize (b) $\beta=3\times 10^{-4}$}
\end{minipage}
\begin{minipage}[!t]{0.32\linewidth}
\centering
 \includegraphics[scale=0.08]{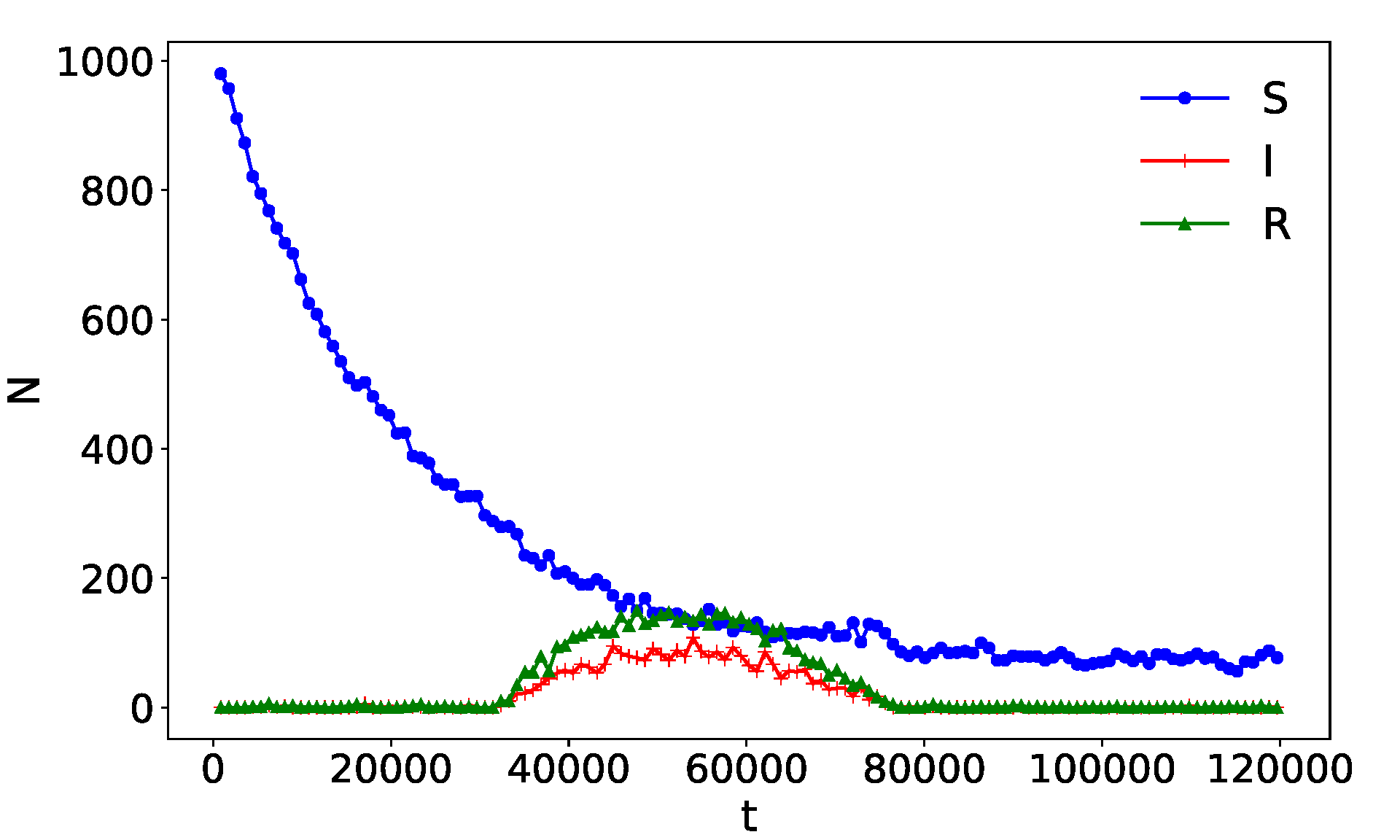}
\parbox{7cm}\centering{\footnotesize (c) $\beta=4\times 10^{-4}$}
\end{minipage}
\begin{minipage}[!t]{0.32\linewidth}
\centering
 \includegraphics[scale=0.08]{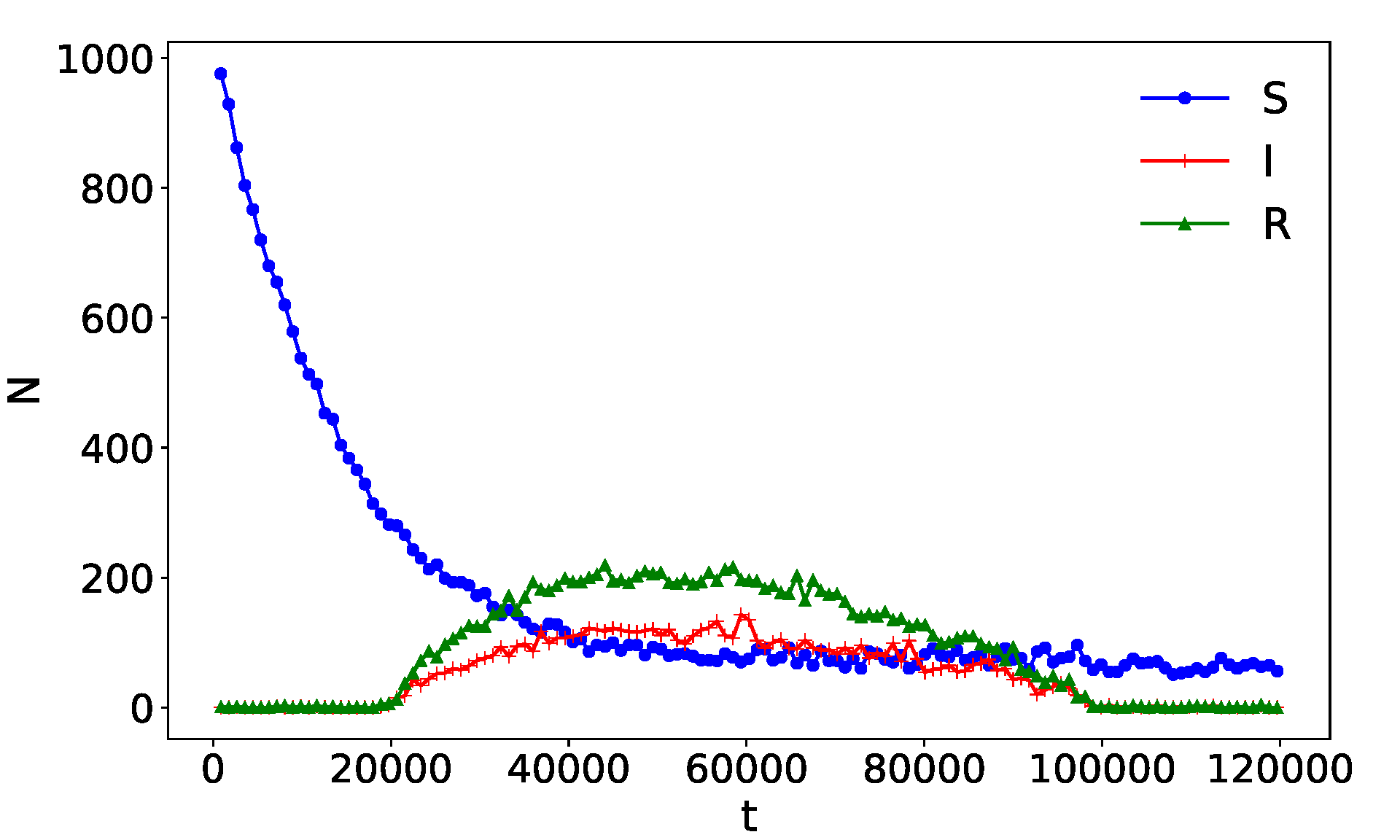}
\parbox{4cm}\centering{\footnotesize (d) $\beta=6\times 10^{-4}$}
\end{minipage}
\begin{minipage}[!t]{0.32\linewidth}
\centering
 \includegraphics[scale=0.08]{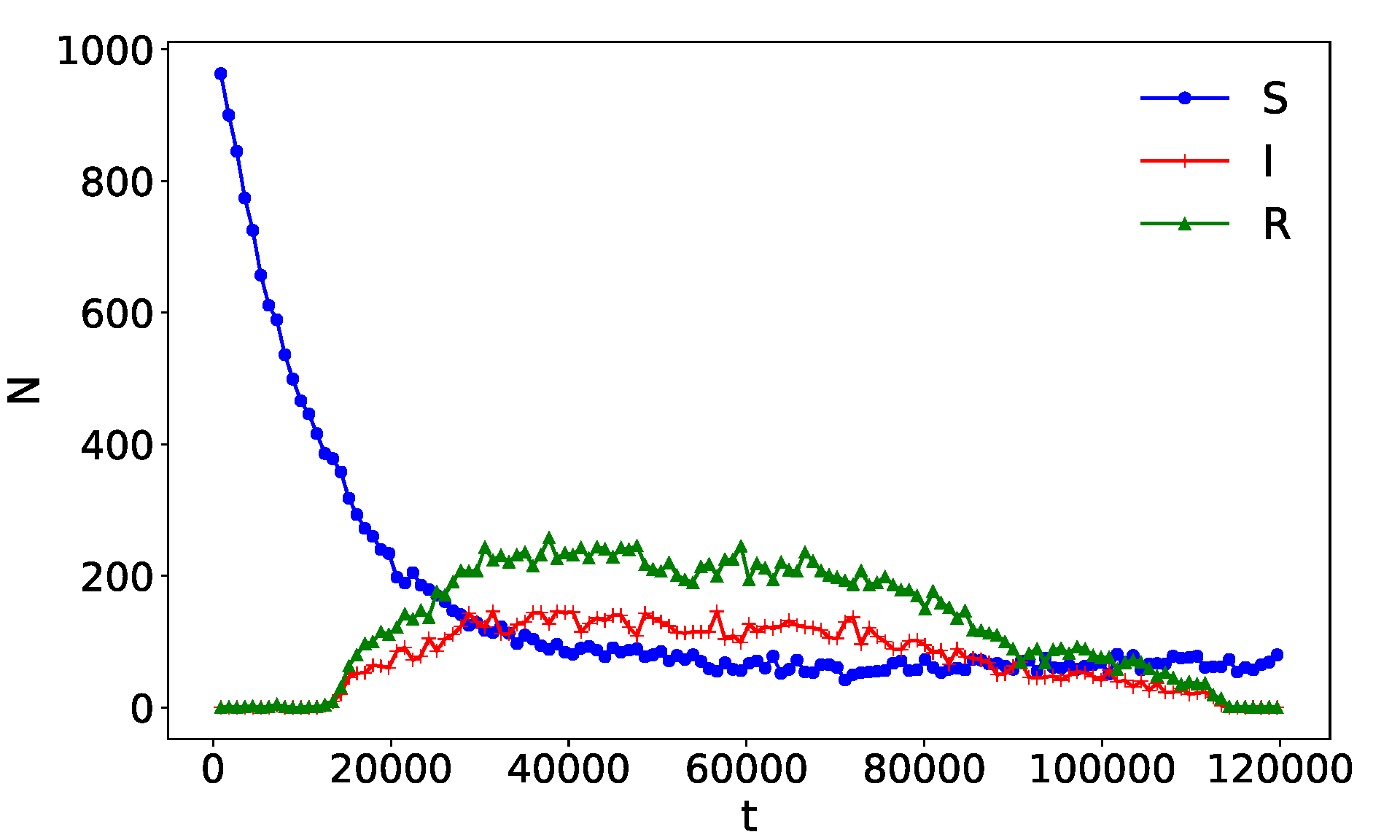}
\parbox{4cm}\centering{\footnotesize (e) $\beta=7\times 10^{-4}$}
\end{minipage}
\begin{minipage}[!t]{0.32\linewidth}
\centering
 \includegraphics[scale=0.08]{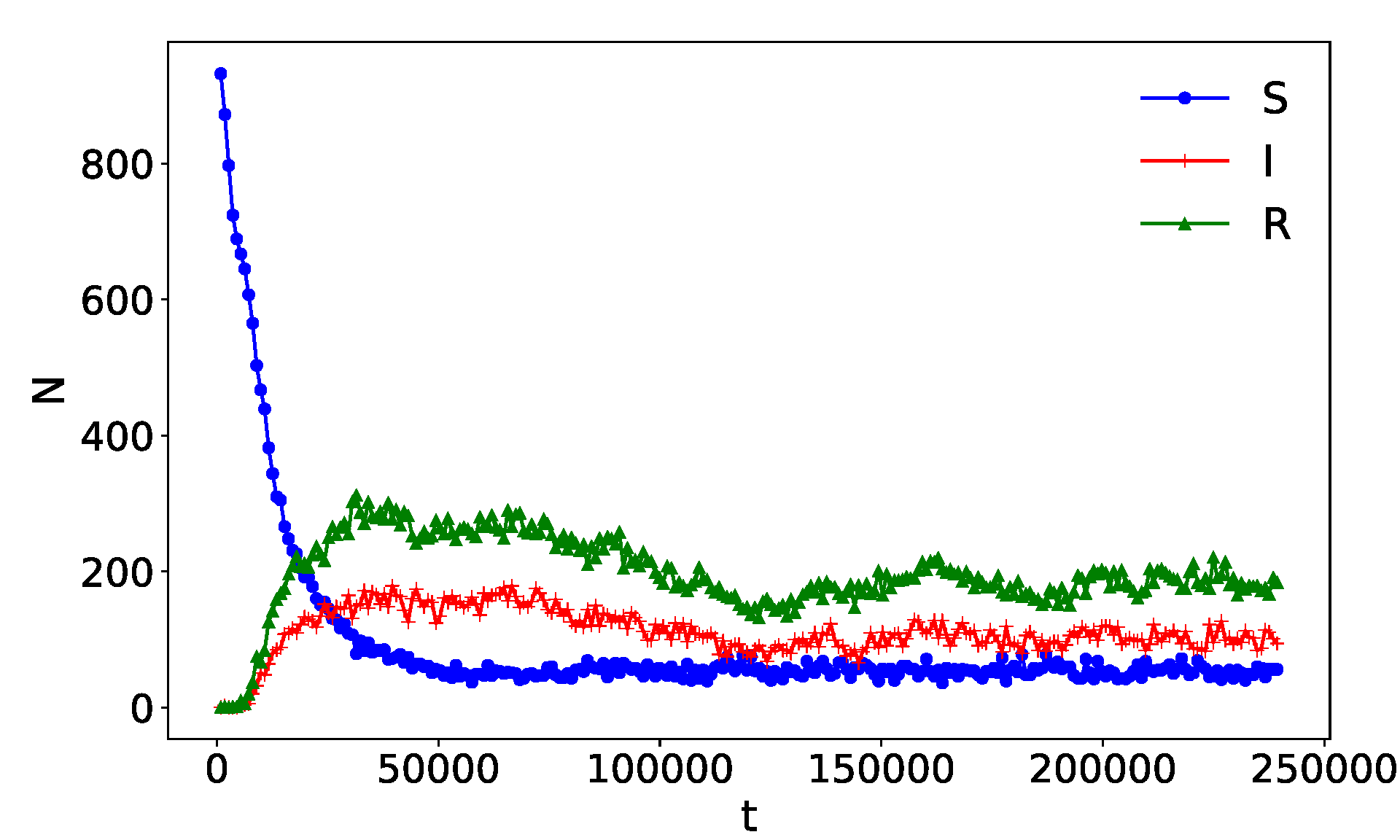}
\parbox{4cm}\centering{\footnotesize (f) $\beta=8\times 10^{-4}$}
\end{minipage}
\caption{\label{fig:yuzhi}The population size of the three epidemic states $S$, $I$, and $R$ varying with time with different values of the infected rate $\beta$. $\beta$ is respectively set as $2\times 10^{-4}$, $3\times 10^{-4}$, $4\times 10^{-4}$, $6\times 10^{-4}$, $7\times 10^{-4}$, $8\times 10^{-4}$ in (a)-(f). The termination time is set long enough as $t$=$1.2\times 10^5$ in (a)-(e) and $t$=$2.4\times 10^5$ in (f). Within the observation time, in (a)-(e), the infected population all experience an observable growth and then turn to 0 or a little above 0, in which situation the epidemic can only spread within a local space and then dies out. In (f), the infected population size is still over 150 within $t$=$1.2\times 10^5$, however, it also eventually turns stationary given a longer time.}
\end{figure*}

\begin{figure*}[htbp]
\centering

\begin{minipage}[t]{0.25\linewidth}
\centering
\includegraphics[scale=0.11]{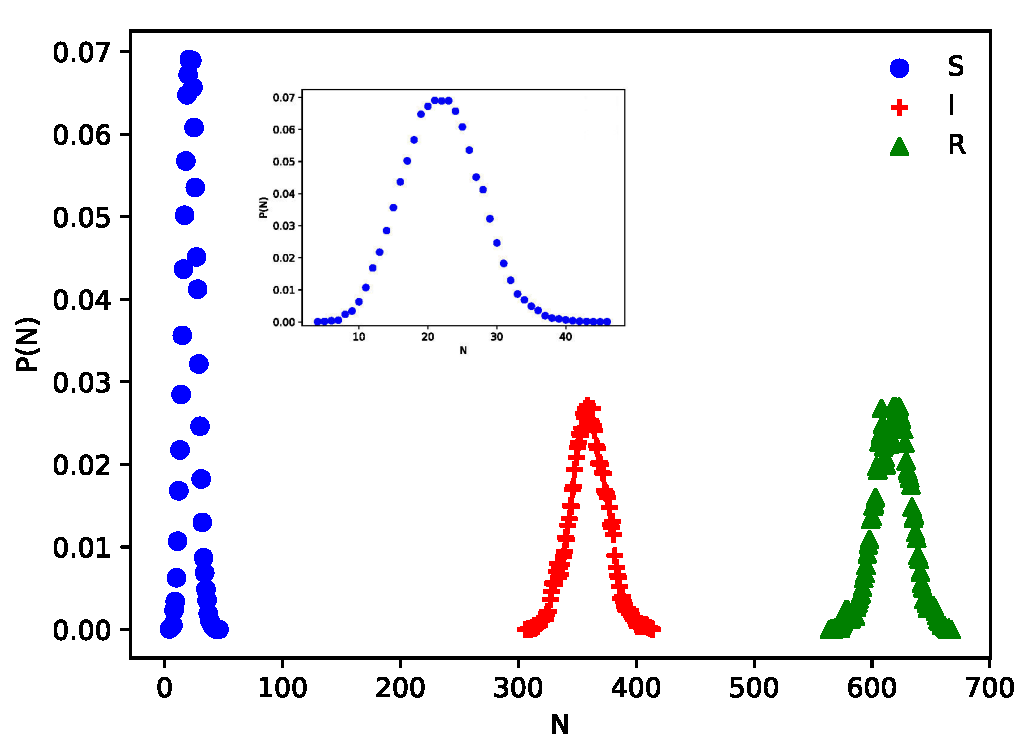}
\parbox{5cm}\centering{\footnotesize \hspace{0.5cm}(a) Initial parameter setting}
\end{minipage}%
\begin{minipage}[t]{0.25\linewidth}
\centering
\includegraphics[scale=0.11]{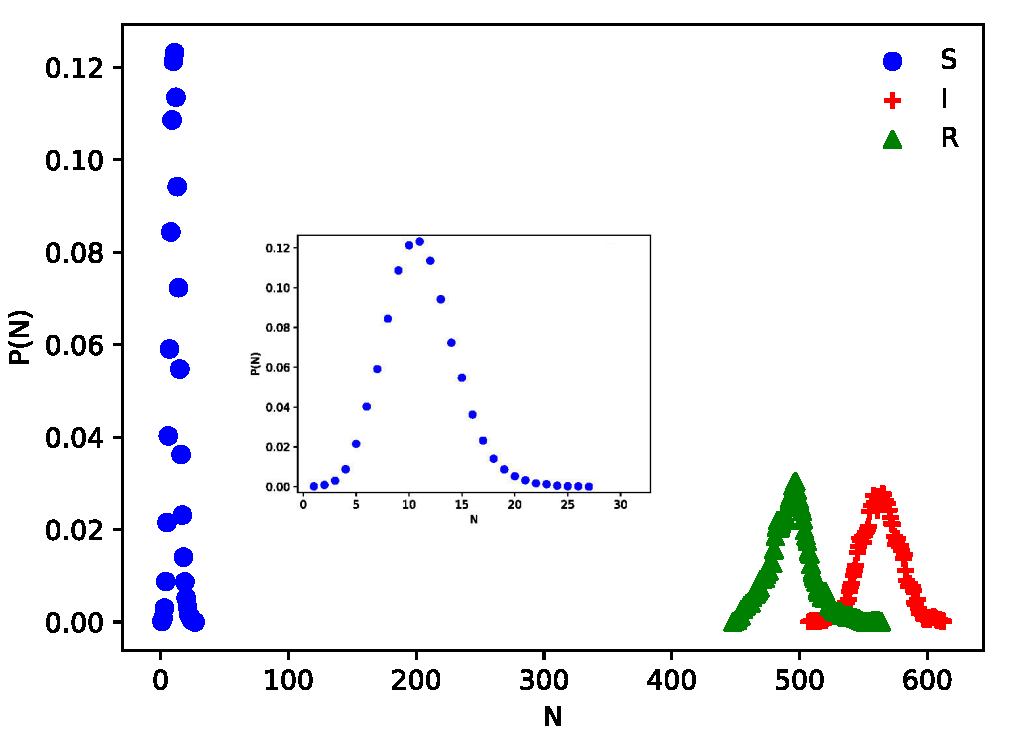}
\parbox{5cm}\centering{\footnotesize \hspace{0.5cm}(b) $\gamma$=0.35}
\end{minipage}%
\begin{minipage}[t]{0.25\linewidth}
\centering
\includegraphics[scale=0.11]{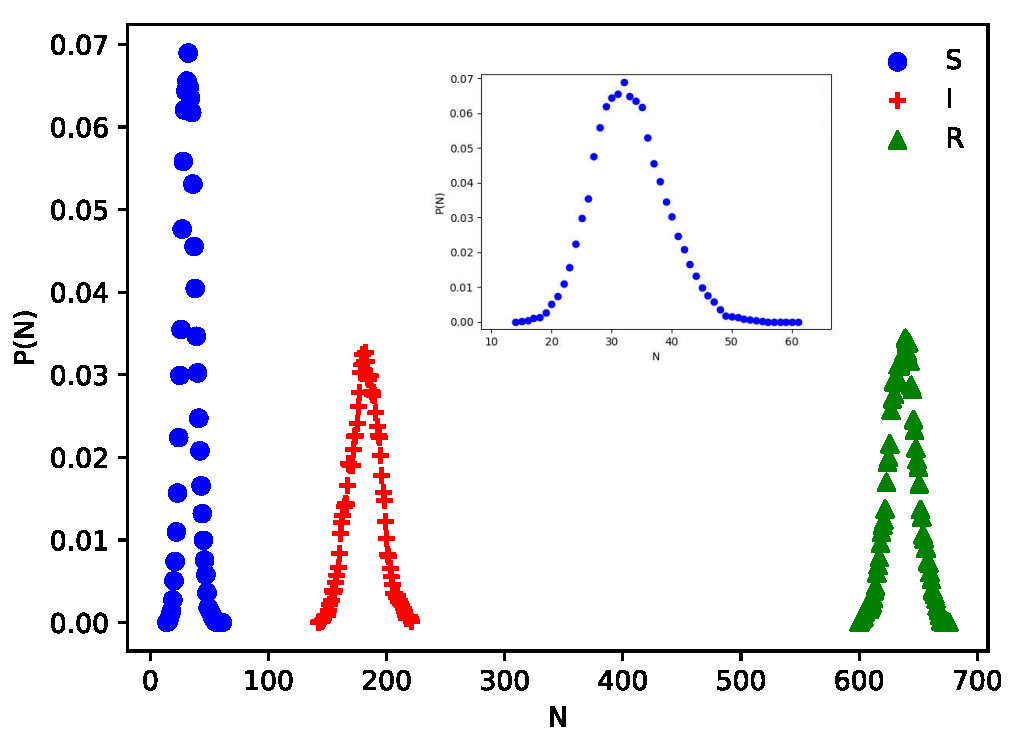}
\parbox{5cm}\centering{\footnotesize \hspace{0.5cm}(c) $\gamma$=1.4}
\end{minipage}%
\begin{minipage}[t]{0.25\linewidth}
\centering
\includegraphics[scale=0.11]{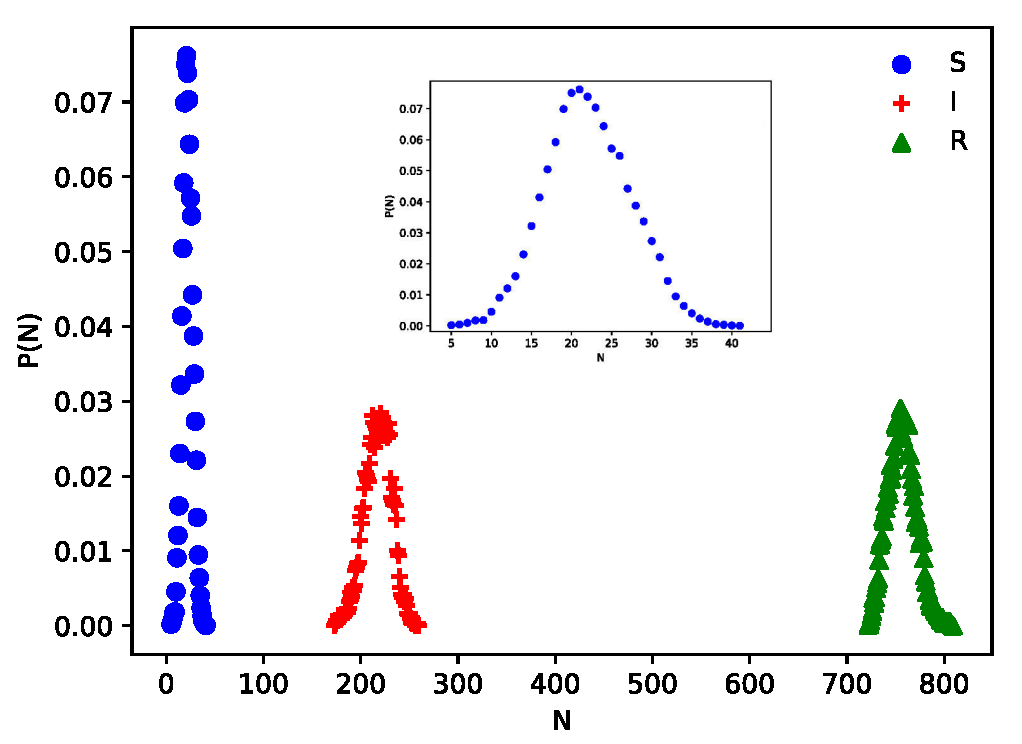}
\parbox{5cm}\centering{\footnotesize \hspace{0.5cm}(d) $\alpha$=0.2}
\end{minipage}%

\begin{minipage}[t]{0.25\linewidth}
\centering
\includegraphics[scale=0.1]{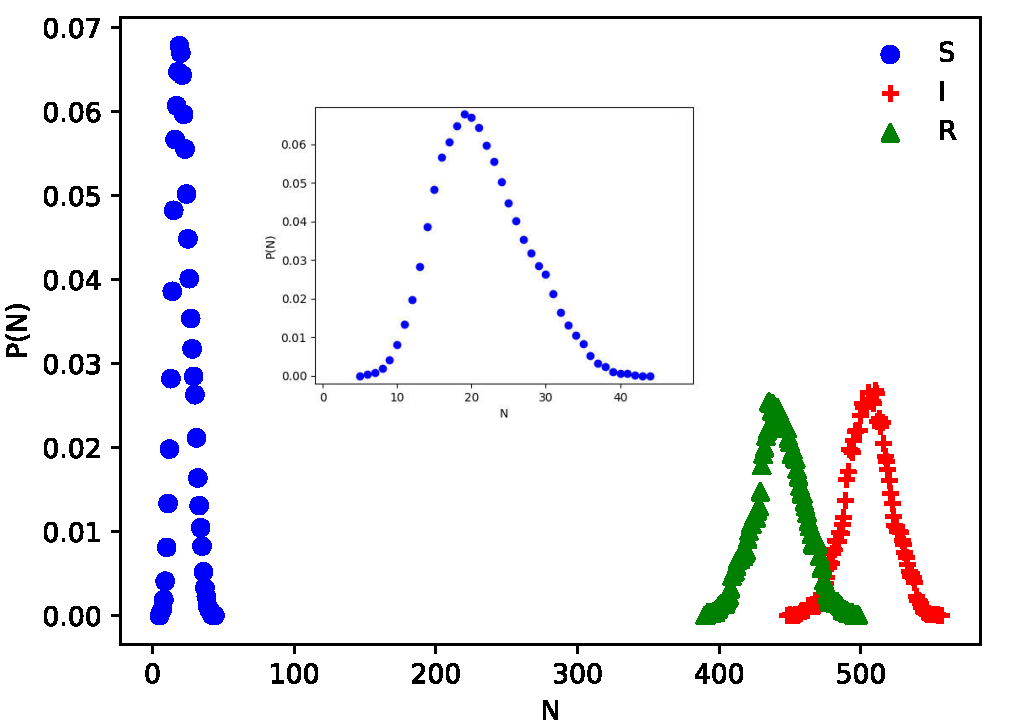}
\parbox{5cm}\centering{\footnotesize \hspace{0.5cm}(e) $\alpha$=0.8}
\end{minipage}%
\begin{minipage}[t]{0.25\linewidth}
\centering
\includegraphics[scale=0.1]{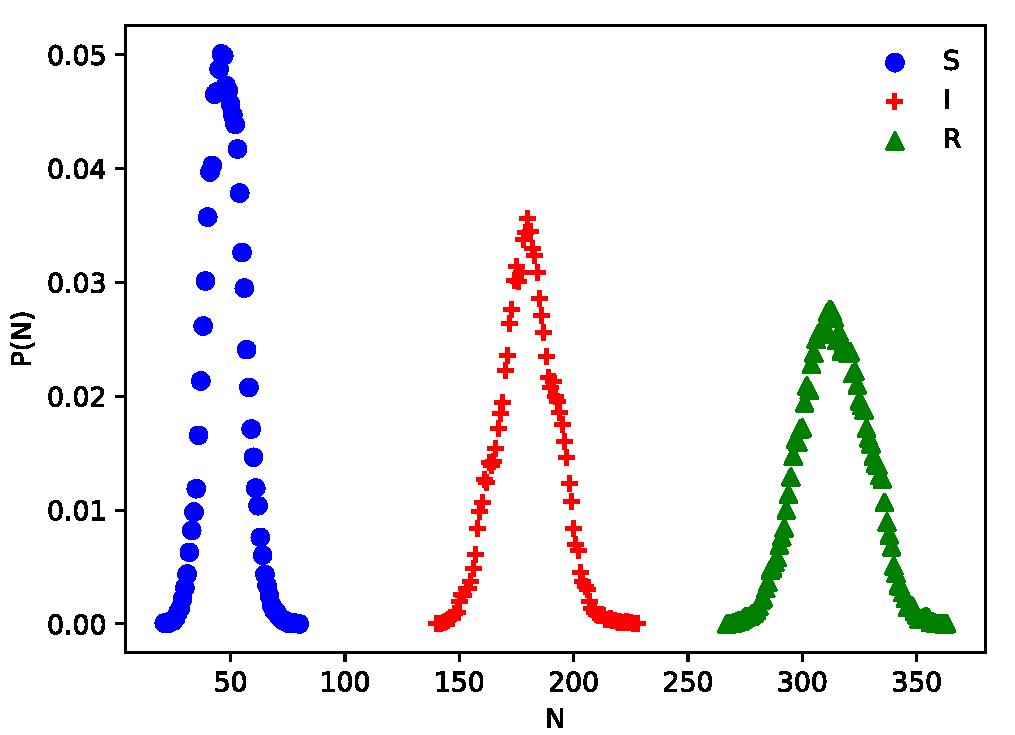}
\parbox{5cm}\centering{\footnotesize \hspace{0.5cm}(f) $\beta$=0.0025}
\end{minipage}%
\begin{minipage}[t]{0.25\linewidth}
\centering
\includegraphics[scale=0.1]{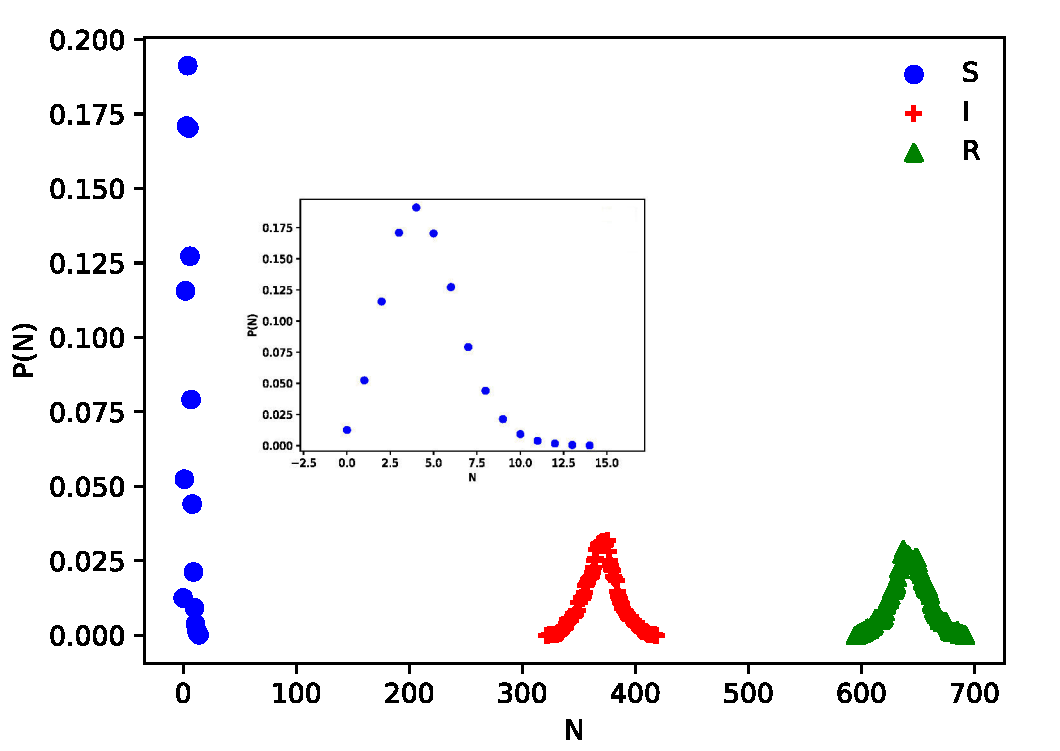}
\parbox{5cm}\centering{\footnotesize \hspace{0.5cm}(g) $\beta$=0.02}
\end{minipage}%

\centering
\caption{\label{fig:ds}Stationary distributions of the population size of three epidemic state $S$, $I$, and $R$, represented by blue, red crosses, and green scatters. The distributions in (a)-(g) subsequently correspond to the population size sequence under the stationary situation in Fig. \ref{fig:n123} (a), (b), (d), (e), (h), and (g). The population size distributions of $S$, $I$, and $R$ with different parameters all demonstrate Poisson distributions, and the point with the largest probability in each distribution is relevant to the stationary value of the population size in Fig. \ref{fig:n123}.}
\end{figure*}

\begin{table*}[htbp]
  \centering
  \caption{Expectation of population sizes of state $S$, $I$, and $R$ with different parameters}
    \begin{tabular}{cccccccc}
    \hline
    \hline
          & Benchmark & $\gamma$=0.35 & $\gamma$=1.4 & $\alpha$=0.2 & $\alpha$=0.8 & $\beta$=0.0025 & $\beta$=0.02 \\
\cline{2-8}    E[S]  & 19    & 11    & 33    & 22    & 21    & 48   & 5 \\
    E[I]  & 354   & 498   & 182   & 207   & 505   & 387   & 178 \\
    E[R]  & 614   & 572   & 637   & 741   & 443   & 678   & 314 \\
    \hline
    \hline
    \end{tabular}%
  \label{tab:e}%
\end{table*}%

As we see from Figs. \ref{fig:yuzhi} (a)-(f), the different values of $\beta$ have different descend rates, and the susceptible population size decreases faster with higher $\beta$. In detail, in Fig. \ref{fig:yuzhi} (a), the population size $S$ with $\beta$=$2\times 10^{-4}$ does not reach stationarity when $t$=$1.2\times 10^5$, while that with $\beta$=$8\times 10^{-4}$ has reached stationarity when $t$=5000 (Fig. \ref{fig:yuzhi} (f)). The infected and recovered population with different $\beta$s in Figs. \ref{fig:yuzhi} (a)-(e) all grow and then decrease to 0 or a little above 0 within $t$=$1.2\times 10^5$. Especially, in Figs. \ref{fig:yuzhi} (a)-(c), there is less than one-third of the whole observation period during which the infected population grows obviously. This indicates that the epidemic only spreads around within a comparatively short time and then dies out. For the rest of the time, the disease can only spread within a local region not globally. Besides, the duration when there is an obvious population size $I$ gets long with $\beta$ increasing, which suggests that higher $\beta$ raises the duration time of epidemic spread around. For $\beta$=$8\times 10^{-4}$, the situation is different for the infected population size presented in Fig. \ref{fig:yuzhi} (f). The infected and recovered population sizes do not converge to 0 while experiencing comparative great fluctuations after $t$=$1.25\times 10^{5}$, at 100 ($I$) and 200 ($R$). In this case, the epidemic can spread around globally. Based on the above simulations, we find that with $\beta$ rising, the epidemic spreads from locally to globally.

\subsection{Stationary Distributions of Population Size of Different States}
Based on the results in the last sub-section, the population size of the three epidemic states can be stationary, and fluctuating at a certain value under restraint conditions for the inflow and outflow rate. For further study, we next investigate stationary distributions of population sizes of different states.

As is illustrated in Fig. \ref{fig:ds}, distributions of the $S$, $I$, and $R$ population size are plotted by blue circles, red crosses, and green triangles respectively. The distribution in Figs. \ref{fig:ds} (a)-(g) orderly corresponds to the population sizes in Figs. \ref{fig:n123} (a), (b), (d), (e), (h), and (g). From the shape of these distributions, we find that they all follow Poisson distributions. This is because the population size of each state fluctuates at a specific value which commands the feature of Poisson distributions.

In Fig. \ref{fig:ds} (a), from the blue plot, the population size of state $S$ with the highest probability is 20 approximately. From the red plot, the $I$ population size with the highest probability is about 350, and from the green plot, the $R$ population size is about 600. These results correspond to the situation in Fig. \ref{fig:n123} (a) where $S$, $I$, and $R$ population respectively fluctuates at 20, 350, and 600 after they are stationary. A similar situation is also shown in Figs. \ref{fig:ds} (b)-(g), where the value with the highest probability in distributions is the value at which population sizes fluctuate. Besides, compared to the distributions in Fig. \ref{fig:ds} (a), in Fig. \ref{fig:ds} (b), there is a left shift of the distribution of $S$ and $R$ population size, and the distribution $I$ population size moves right, suggesting that a lower recovered rate $\gamma$ leads to less susceptible and infected individuals while more recovered individuals. In contrast, in Fig. \ref{fig:ds} (c), there is a left shift of the distribution of $I$ population size and a right shift for the distribution of $S$ and $R$ population size.

From Figs. \ref{fig:ds} (d) and (e), the $S$ population size does not change as $\alpha$ increases or decreases, while the distribution of the $I$ population size moves left and that of the $R$ population size moves right, which indicates that a lower reviving rate $\alpha$ leading to less infected individuals while more recovered individuals. In Fig. \ref{fig:ds} (f), with $\beta$ decreasing, the distribution of the $S$ population size moves right, while that of the $I$ and $R$ population size moves left. In Fig. \ref{fig:ds} (g) where $\beta$ increases, there is a small right shift for the distributions of the $I$ and $R$ population size, and a left shift for $S$. This suggests that a lower infected rate $\beta$ leads to a larger susceptible population while a smaller infected and recovered population.
Then, we calculate the expectation of the population sizes with different parameters, which is demonstrated in Table. \ref{tab:e}. We see that the expectation values are in accordance with their distribution plots.

\begin{figure*}[htbp]
\begin{minipage}[!t]{0.32\linewidth}
\centering
 \includegraphics[scale=0.083]{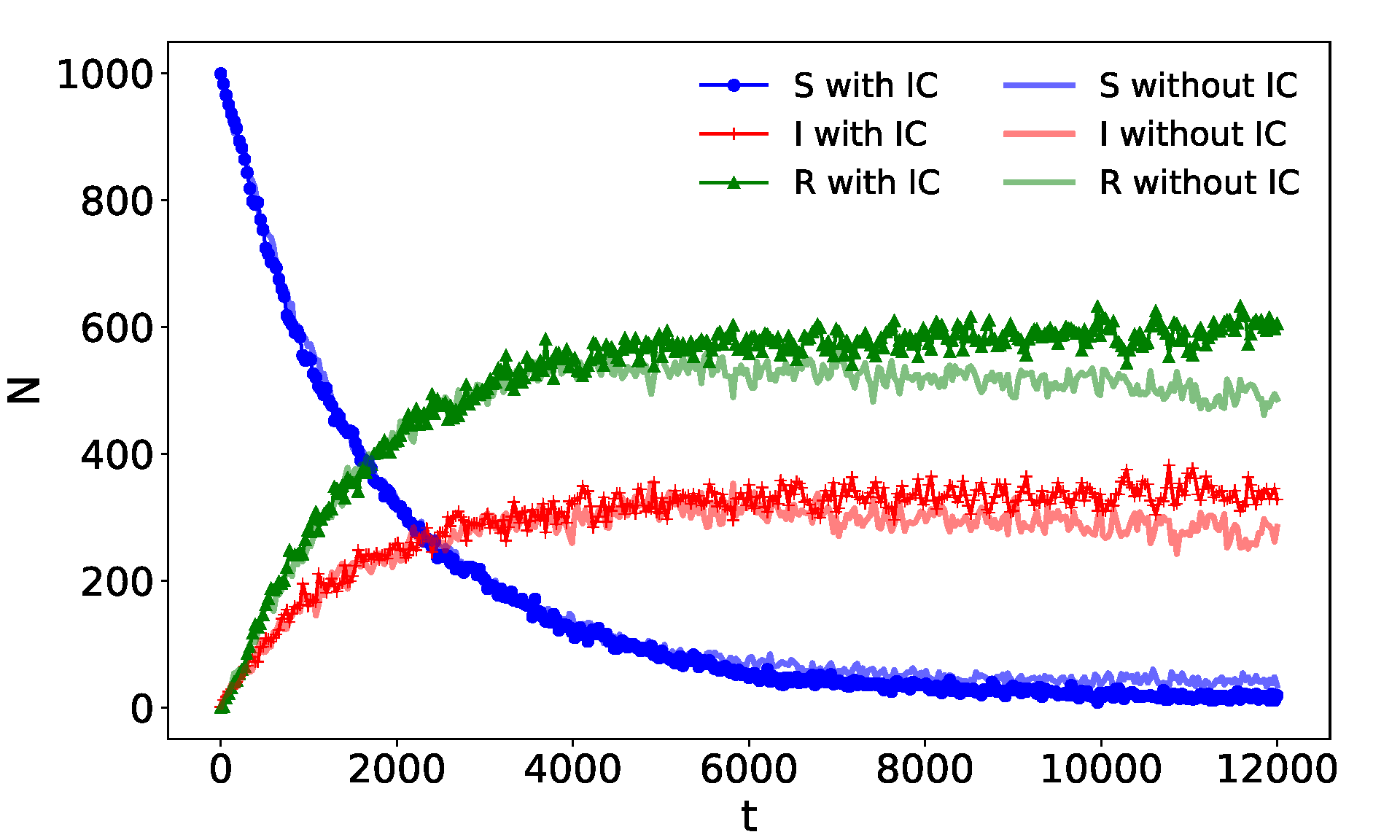}
\parbox{4.4cm}\centering{\footnotesize \hspace{0.1cm}(a) $\lambda=1.5$, $\mu=0.025$}
\end{minipage}
\begin{minipage}[!t]{0.32\linewidth}
\centering
 \includegraphics[scale=0.083]{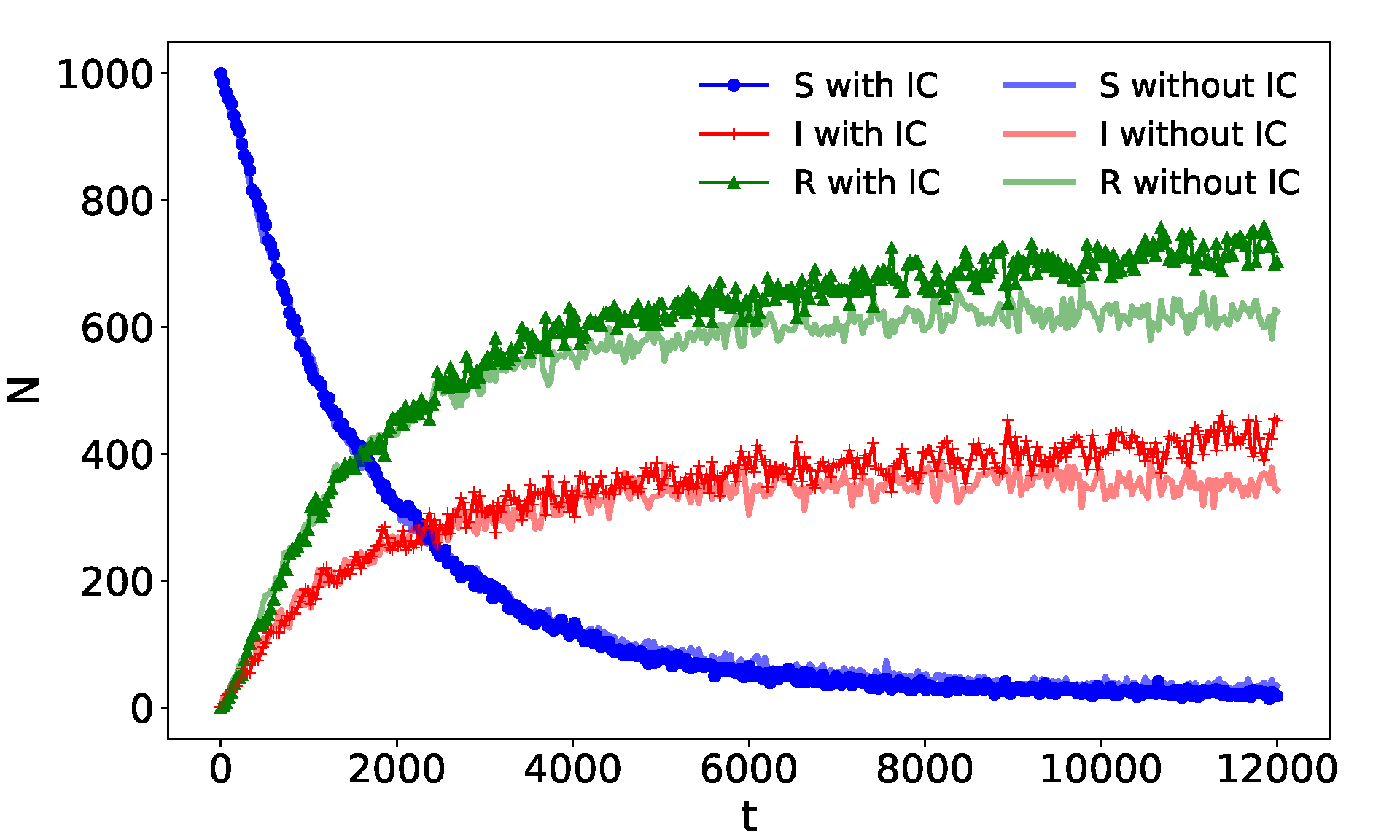}
\parbox{4.4cm}\centering{\footnotesize \hspace{0.1cm}(b) $\lambda=1.5$, $\mu=0.017$}
\end{minipage}
\begin{minipage}[!t]{0.32\linewidth}
\centering
 \includegraphics[scale=0.083]{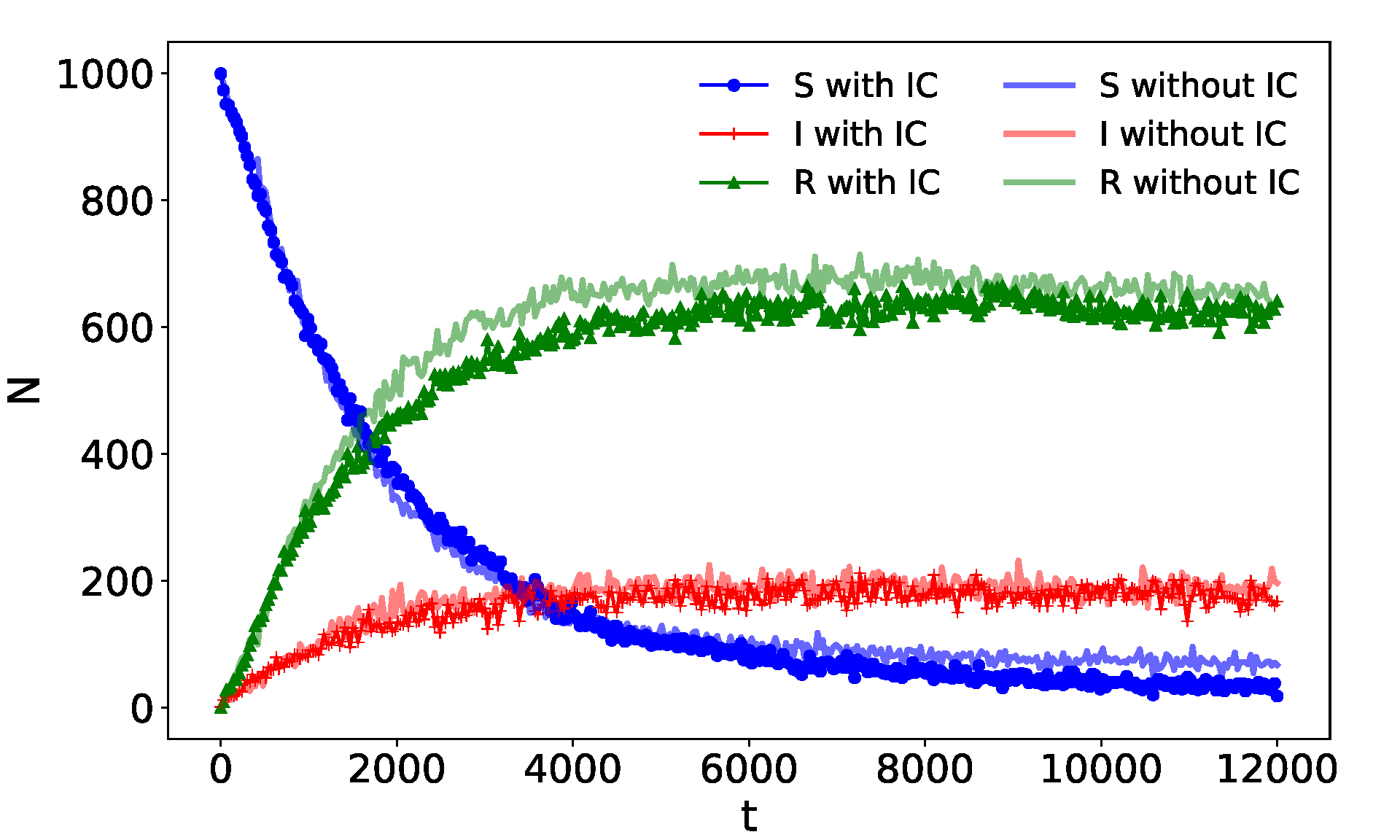}
\parbox{8.5cm}\centering{\footnotesize\hspace{0.5cm}\vspace{0.15cm}{(c) $\gamma=0.14$}}
\end{minipage}
\begin{minipage}[!t]{0.32\linewidth}
\centering
 \includegraphics[scale=0.083]{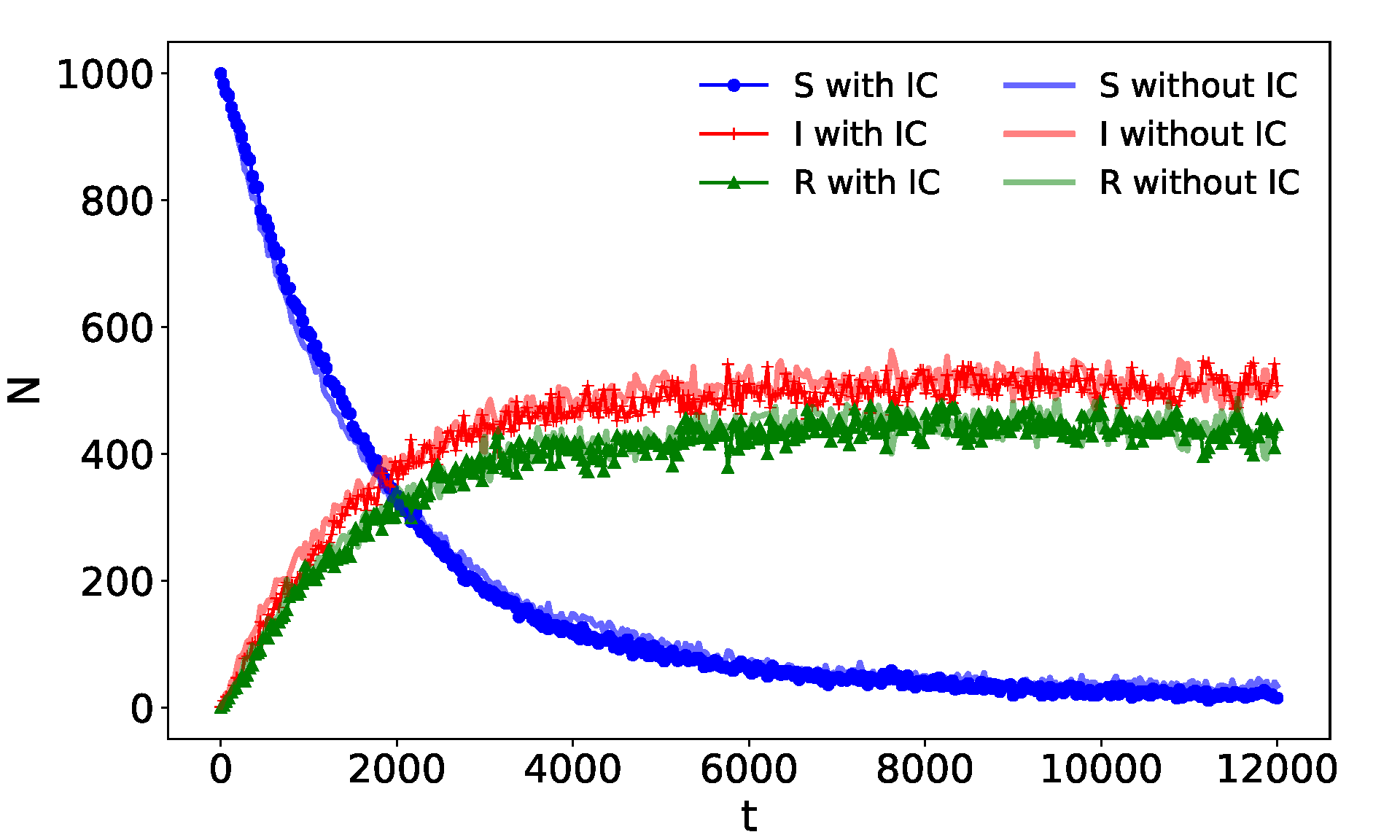}
\parbox{8.5cm}\centering{\footnotesize \hspace{0.5cm}\vspace{0.15cm}{(d) $\alpha=0.8$}}
\end{minipage}3
\begin{minipage}[!t]{0.32\linewidth}
\centering
 \includegraphics[scale=0.083]{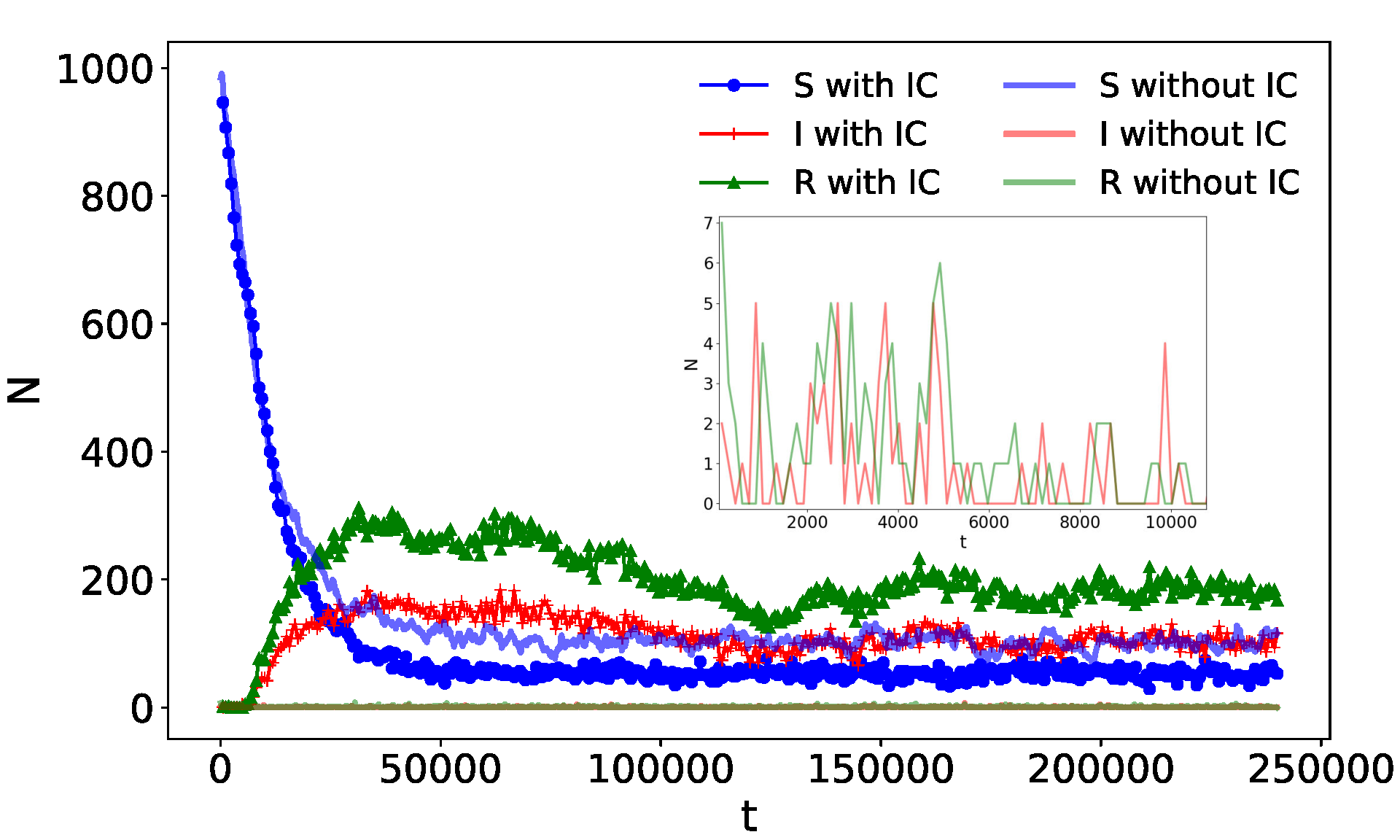}
\parbox{4.4cm}\centering{\footnotesize \hspace{0.1cm}(e) $\beta=0.0008$}
\end{minipage}
\begin{minipage}[!t]{0.32\linewidth}
\centering
 \includegraphics[scale=0.083]{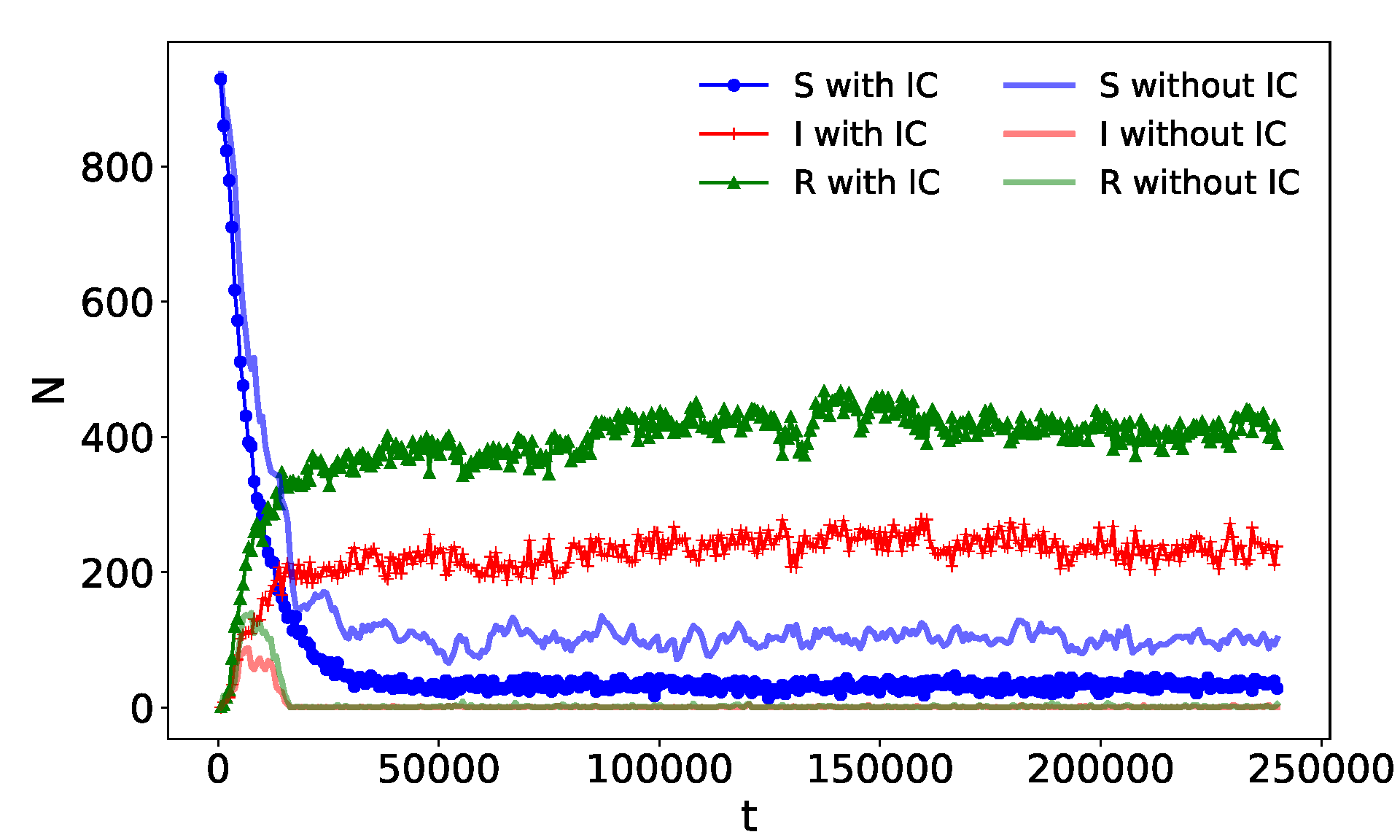}
\parbox{8.5cm}\centering{\footnotesize \hspace{0.5cm}\vspace{0.15cm}{(f) $\beta=0.0012$}}
\end{minipage}
\caption{\label{fig:comp}The comparison of the epidemic spreading with and without indirect contacts given different settings of parameters. Dark marked lines and light non-marked lines respectively represent the population sizes with indirect contacts and without indirect contacts. The initial parameters are set in (a), where population sizes with indirect contacts reach stationary, while the population sizes without indirect contacts are not stationary under this pair of $\lambda$ and $\mu$. (b) shows the opposite situation compared to (a), where the population sizes without indirect contacts are stationary with $\lambda$=1.5 and $\mu$=0.017. (c) and (d) present that the population sizes of three states varying with time are the same under the same $\gamma$ and $\alpha$. (e) and (f) display the situation with different $\beta$s, and the inserted plot in (e) displays the infected population size without indirect contacts.}
\end{figure*}
\subsection{Effect of indirect contacts on Epidemic Spreading Process}
Next, we compare the epidemic spreading with indirect contacts (denoted by IC in Fig. \ref{fig:comp}) and without indirect contacts. The initial setting is the same as that in Sec. \ref{sec:A}. Based on the initial setting, we adjust five parameters orderly, and the population sizes of the three epidemic states are presented in Fig. \ref{fig:comp}.

From Fig. \ref{fig:comp} (a), the population sizes of $S$ are overlapped under the model with and without indirect contacts. The population size of $I$ and $R$ without indirect contacts slowly increase after $t$=5000, by contrast, those with indirect contacts are stationary. In Fig. \ref{fig:comp} (b), we lower the outflow rate $\mu$, and the population size of $I$ and $R$ without indirect contacts are stationary with $\mu$=0.017 while those with indirect contacts have an upwards trend. The results in Figs. \ref{fig:comp} (a) and (b) indicate that the indirect contact mechanism has an influence on the stationary condition for the inflow and outflow rate, in particular, indirect contacts enhance the value of the outflow rate for the stationary situation given a specific inflow rate. In Figs. \ref{fig:comp} (c) and (d), under each model's stationary conditions for $\lambda$ and $\mu$, we adjust $\gamma$ and $\alpha$, and the marked lines and non-marked lines are overlapped, indicating the population sizes of three states varying with time are the same. In detail, in Fig. \ref{fig:comp} (c), the population sizes of $I$ with and without indirect contacts both increase as $\gamma$ increases, while the population sizes of $R$ with and without indirect contacts both decrease. A similar conclusion can be obtained in Fig. \ref{fig:comp} (d).

Then, we set $\beta$=0.0008 and 0.0012 respectively in Figs. \ref{fig:comp} (e) and (f), and $\lambda$ and $\mu$ are set as the value under their respective conditions which are the same as that in Figs. \ref{fig:comp} (a) and (b). From Fig. \ref{fig:comp} (e), the population size of $I$ and $R$ with indirect contacts reach stationarity, fluctuating at 100 and 200 when $\beta$=0.0008, while that without indirect contacts keep 0 or a little above 0. In Fig. \ref{fig:comp} (f), the population size of $I$ and $R$ with indirect contacts fluctuates at 200 and 400 higher than that with $\beta$=0.0012. The population size of $I$ and $R$ without indirect contacts experience growth to 80 and 140 and then declines to 0 or a little above 0 in a comparatively short time. The results in Figs. \ref{fig:comp} (e) and (f) show that the infected population size with indirect contacts is larger than that without indirect contacts under the same infected rate $\beta$. Both situations in Figs. \ref{fig:comp} (e) and (f) indicate that the epidemic only spread locally with indirect contacts, while the epidemic would spread globally leading to a certain amount of infected population existing in a quite long time without indirect contacts. Therefore, indirect contacts caused by the network evolution enhance epidemic spreading on networks.

\subsection{Validation of reproductive number}
\begin{figure}[t]
    \centering
    \includegraphics[width=0.5\textwidth]{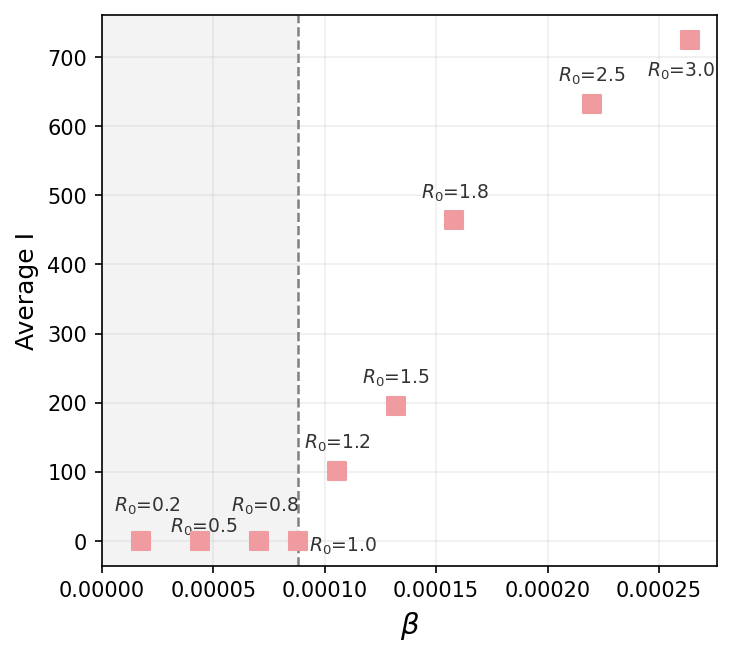}
    \caption{The average number of infected individuals of 10 simulations given different values of the infected rate $\beta$. The dashed vertical line indicates the theoretical threshold of $\beta$ ($\beta_C=0.000088$) given a set of parameters. On the x-axis, the values of $\beta$ is respectively set as $0.2\beta_C$, $0.5\beta_C$, $0.8\beta_C$, $\beta_C$, $1.2\beta_C$, $1.5\beta_C$, $2\beta_C$, $2.5\beta_C$, and $3\beta_C$. Below the threshold, the number of infected individuals equals to 0, which indicates that the disease dies out, while above it, the number of infected individuals is larger than 0, which indicates an epidemic.}
    \label{fig:threshold}
\end{figure}

To validate the theoretical reproduction number derived in Thm. \ref{th:2}, we calculate the theoretical epidemic threshold based on the reproduction number and conduct simulations with infection rates below, equal to, and above this threshold to examine whether an outbreak occurs. In detail,  we set the parameters as follows: $\lambda=45$, $\alpha = 0.4$, $\gamma = 0.7$, $\mu = 0.05$, $\pi_S^{in} = 0.85$, $\pi_R^{in} = 0.15$, $\pi_S^{out} = 0.90$, $\pi_I^{out} = 0.05$, $\pi_R^{out} = 0.05$. The initial number of infected individuals is set as $10$. As assumed in Thm. \ref{th:2}, we set $\pi_I^{in} = 0$, since persistent infections would occur when $\pi_I^{in} > 0$. The initial setting of the network is the same as that described in section 3.1. We compute the final number of infected individuals under different values of $\beta$ (set as $0.2\beta_C$, $0.5\beta_C$, $0.8\beta_C$, $\beta_C$, $1.2\beta_C$, $1.5\beta_C$, $2\beta_C$, $2.5\beta_C$, and $3\beta_C$). For each parameter setting, we perform 10 stochastic simulations and report the average final number of infected individuals. The results are shown in Fig. \ref{fig:threshold}. The theoretical threshold $\beta_c$ is $0.00088$ marked by the vertical line in the plot. The number of infected individuals is equal to 0 with the values of $\beta$ below or equal to $\beta_c$, while when $\beta$ is above the threshold, a positive number of infections is observed.

\section{Conclusion and Outlook}\label{sec:V}
We model epidemic propagation on the proposed birth-death evolving network with the heritable disconnection mechanism, considering the co-evolution of the epidemic propagation and the network on a unified time scale. The model include both the population migration and indirect contacts into consideration during an epidemic. Besides, we use stochastic methods to analyze the epidemic spreading by constructing a Markovian queueing network. This provides a new perspective for studying epidemic propagation, based on which we can analyze the population sizes of each epidemic state via drawing transition probabilities. Furthermore, in simulations, we have investigated the impact of population migration and indirect contacts on population sizes of the three states and found that the indirect contacts can lower the epidemic threshold for the infected rate and raise the infected population size under stationarity, facilitating epidemic spreading.

We acknowledge the limitations of our work. First, while we propose how the numbers of individuals of the three epidemic states change with time, we do not perform the analyses for the stationary distribution of them. The epidemic dynamics in the system follow a non-homogeneous Markov process, in which the transition probabilities vary with time. It is challenging to derive explicit expressions for the stationary distributions. To this end, we conduct numerical simulations to characterize how the key parameters and indirect contacts affect the epidemic spreading. Nevertheless, we recognize the importance of more rigorous analyses of dynamics on our evolving network. we plan to work on this aspect in our future work.
Besides, we simplify the selection criteria for the inheritor node by selecting randomly, which should be different in our original network model. We will work on these limitations in future work.

\section{Acknowledgements}
The authors thank the support from Natural Science Foundation of Chongqing under Grant CSTB2025YITP-QCRCX0007.

\printcredits

\bibliographystyle{cas-model2-names}
\bibliography{cas-refs}

\end{document}